\documentclass[aps,prd,preprint,superscriptaddress,showpacs]{revtex4}

\usepackage{graphicx}

\def\bwt{\begin{widetext}}

\def\ewt{\end{widetext}}

\def\be{\begin{equation}}

\def\ee{\end{equation}}

\def\bea{\begin{eqnarray}}

\def\eea{\end{eqnarray}}

\def\bean{\begin{eqnarray*}}

\def\eean{\end{eqnarray*}}

\def\bary{\begin{array}}

\def\eary{\end{array}}

\def\bit{\begin{itemize}}

\def\eit{\end{itemize}}

\def\su5u1{SU(5) \times U(1)}

\def\fsu5u1{SU(5) \times U(1)'}

\def\so10{SO(10)}

\def\sq20{SO(10) \times SO(10)}

\usepackage[centertags]{amsmath}

\usepackage{amssymb}

\newcommand{\Z}{{\mathbb Z}}

\begin{document}

\title{Flipped and Unflipped $SU(5)$ as Type IIA Flux Vacua}

\author{Ching-Ming Chen}

\affiliation{George P. and Cynthia W. Mitchell Institute for
Fundamental Physics, Texas A$\&$M University, College Station, TX
77843, USA }

\author{Tianjun Li}

\affiliation{Department of Physics and Astronomy, Rutgers
University, Piscataway, NJ 08854, USA }

\affiliation{Institute of Theoretical Physics, Chinese Academy of
Sciences, Beijing 100080, P. R. China }

\author{Dimitri V. Nanopoulos}

\affiliation{George P. and Cynthia W. Mitchell Institute for
Fundamental Physics, Texas A$\&$M University, College Station, TX
77843, USA }

\affiliation{Astroparticle Physics Group, Houston Advanced
Research Center (HARC), Mitchell Campus, Woodlands, TX 77381, USA}

\affiliation{Academy of Athens, Division of Natural Sciences, 28
Panepistimiou Avenue, Athens 10679, Greece }

\date{\today}

\begin{abstract}

On Type IIA orientifolds with flux compactifications in
supersymmetric AdS vacua, we for the first time construct 
 $SU(5)$ models with three anti-symmetric ${\bf 10}$ representations
and without symmetric ${\bf 15}$ representations.
We show that all the pairs of the anti-fundamental ${\bf \bar 5}$ and 
fundamental ${\bf 5}$ representations can obtain
 GUT/string-scale vector-like masses after the additional
gauge symmetry breaking via supersymmetry preserving Higgs
mechanism. Then we have exact three ${\bf \bar 5}$, and
no other chiral exotic particles that are charged under
$SU(5)$  due to the non-abelian anomaly free condition. 
Moreover, we can break the $SU(5)$ gauge symmetry 
 down to the SM gauge symmetry via D6-brane
splitting, and solve the doublet-triplet splitting problem.
Assuming that the extra one (or several) pair(s) of Higgs doublets and
adjoint particles obtain GUT/string-scale masses
via high-dimensional operators, we only have
the MSSM in the observable sector 
below the GUT scale. Then the observed low energy gauge
couplings can be generated via RGE running
if we choose the suitable grand unified gauge coupling by 
 adjusting the string scale.  Furthermore, 
we construct the first flipped $SU(5)$ model with exact 
three ${\bf 10}$, and the first flipped $SU(5)$ model in which
all the Yukawa couplings are allowed by the global $U(1)$
symmetries.

\end{abstract}


\preprint{ACT-03-06, MIFP-06-09, hep-th/0604107}

\maketitle

\section{Introduction}

The major challenge and lasting problem in string phenomenology is
to construct realistic Standard-like string models with moduli
stabilization. In the beginning, string model building was mainly
concentrated on the weakly coupled heterotic string theory, and
rather successful models like flipped $SU(5)$~\cite{Barr:1981qv,FSU(5)N}
in its stringy form were constructed~\cite{AEHN}. Meanwhile, the first
standard compactification of strong coupled heterotic string
theory or M-theory on $S^1/Z_2$ was given in
Ref.~\cite{Li:1997sk}. Due to the advent of  D-branes in the
second string revolution~\cite{JPEW}, we can construct consistent
four-dimensional chiral models with non-abelian gauge symmetry on
Type II orientifolds.

Type II orientifolds with intersecting D-branes have been 
extremely valuable in string model building during 
the last few years. The
chiral fermions can arise from the intersections of D-branes in
the internal space~\cite{bdl} with T-dual description in terms of
magnetized D-branes~\cite{bachas}. In addition, a lot of
non-supersymmetric three-family Standard-like models and grand
unified models without Ramond-Ramond (RR) tadpole on Type IIA
orientifolds with intersecting D6-branes were
constructed~\cite{Blumenhagen:2000wh,Angelantonj:2000hi,Blumenhagen:2005mu}.
However, there generically exist the uncancelled
Neveu-Schwarz-Neveu-Schwarz (NSNS) tadpoles and the gauge
hierarchy problem. To solve these two problems, the first
quasi-realistic supersymmetric models have been
constructed in Type IIA theory on $\mathbf{T^6/(\Z_2\times \Z_2)}$
orientifold with intersecting D6-branes~\cite{CSU}. Subsequently, 
supersymmetric Standard-like models, Pati-Salam models, $SU(5)$
models as well as flipped $SU(5)$ models have been constructed
systematically~\cite{CP,Cvetic:2002pj,CLL,Cvetic:2004nk,Chen:2005ab,Chen:2005mj,Gmeiner:2006vb},
and their  phenomenological consequences have been
studied~\cite{CLS1,CLW}. Also, the supersymmetric constructions on
other orientifolds were discussed as well~\cite{ListSUSYOthers}.
There are two main constraints on supersymmetric model building:
RR tadpole cancellation conditions  and four-dimensional $N=1$
supersymmetry conditions.

However, the moduli stabilization in open string and closed string
sectors is still an open problem, although some of the complex
structure parameters (in the Type IIA picture) and the dilaton field
may be stabilized due to the gaugino condensations in the hidden
sector in some models~\cite{CLW}. Another way to  stabilize the
compactification moduli fields is turning on the supergravity
fluxes~\cite{GVW}. The point is that a supergravity potential can
be generated, and the continuous moduli space of the string vacua
in the four-dimensional effective theory can be lifted. On Type
IIB orientifolds, the supergravity fluxes contribute large
positive D3-brane charges due to the Dirac quantization
conditions, and then modify the global RR tadpole cancellation
conditions significantly and imposes strong constraints on
consistent model building~\cite{CU,BLT}. Thus, one can construct
 three-family and four-family Standard-like
models~\cite{Chen:2005mj,MS,CL,Cvetic:2005bn,Kumar:2005hf,Chen:2005cf}
if and only if one introduces magnetized D9-branes with large
negative D3-brane charges in the hidden and observable sectors. By
the way, it has been recently shown that if non-geometric
fluxes and new Type IIB S-duality fluxes are introduced, 
they can contribute negative D-brane charges to the RR tadpole cancellation
conditions in supersymmetric Minkowski vacua
on Type IIB orientifolds~\cite{Aldazabal:2006up,Villadoro:2006ia}. 
But, we will not consider it in this paper.

The techniques for consistent chiral flux compactifications on
Type IIA orientifolds with intersecting D6-branes were developed
recently~\cite{Grimm:2004ua,Villadoro:2005cu,Camara:2005dc,Camara:2005pr}.
Interestingly enough, in supersymmetric AdS vacua, the metric, NSNS and
RR fluxes can contribute negative D6-brane charges to all the RR
tadpole cancellation conditions, {\it i.~e.}, the RR tadpole
cancellation conditions give no constraints on consistent model
building~\cite{Chen:2006gd}. Thus, the supersymmetric flux models
on Type IIA orientifolds are mainly constrained by
four-dimensional $N=1$ supersymmetry conditions, and then we can
construct rather realistic intersecting D6-brane
models~\cite{Chen:2006gd}.

In this paper, we construct Grand Unified 
Theories (GUTs) such as $SU(5)$ and flipped 
 $SU(5)$ models on Type IIA orientifolds with flux
compactifications in supersymmetric AdS vacua. Although the
up-type quark Yukawa couplings and down-type quark Yukawa
couplings are forbidden respectively in the $SU(5)$ models and
flipped $SU(5)$ models by the $U(1)$ symmetries, these
models do have some interesting features, for example, the gauge
coupling unification which generically can not be realized  in the
other models for D-brane constructions. However, in  the previous
$SU(5)$ model building in Type IIA theory on
$\mathbf{T^6/(\Z_2\times \Z_2)}$ orientifold without fluxes, we
can easily show that we cannot construct models with three
anti-symmetric ${\bf 10}$ representations and without symmetric ${\bf 15}$
representations~\cite{Cvetic:2002pj,Gmeiner:2006vb}. 
In addition, for models with
three anti-symmetric representations and some symmetric
representations, the net number of anti-fundamental ${\bf \bar 5}$
and fundamental ${\bf 5}$ representations  can not be
 three due to the non-abelian anomaly free
conditions, {\it i.~e.}, one does not have exact three families of
the SM fermions (in our convention, we define the net number of
vector-like particles $X$ and $\bar X$ as the number of $X$ minus the
number of $\bar X$ where $X$ can be ${\bf \bar 5}$ or ${\bf 10}$.).
  Moreover, in the previous flipped $SU(5)$
models~\cite{Chen:2005ab,Chen:2005mj,Chen:2005cf}, 
both the net number of ${\bf 10}$ and ${\bf \overline{10}}$
 and the net number of ${\bf \overline{5}}$ and ${\bf {5}}$ are
not three, and at least some Yukawa couplings are 
forbidden by the global $U(1)$ symmetries.

On Type IIA orientifolds with flux compactifications in
supersymmetric AdS vacua, we construct the
 $SU(5)$ models with three anti-symmetric ${\bf 10}$ representations
and without symmetric ${\bf 15}$ representations. Although the net number
of the ${\bf \bar 5}$ and ${\bf 5}$ is three due to the non-Abelian anomaly
free condition, the initial ${\bf \bar 5}$ number $n_{\bf \bar 5}$ is
not three and the initial ${\bf 5}$ number is $n_{\bf \bar 5}-3$. 
We show that all the ${\bf \bar 5}$ and ${\bf 5}$ pairs can obtain
 GUT/string-scale vector-like masses after the extra
gauge symmetry breaking via supersymmetry preserving Higgs
mechanism. So, unlike the previous D-brane models, there
are exact three ${\bf \bar 5}$, and
 no chiral exotic particles that are charged under $SU(5)$.
 In addition, the $SU(5)$ gauge symmetry can be broken down
to the Standard Model (SM) gauge symmetry via D6-brane splitting, and
the doublet-triplet splitting problem can be solved.
If the extra one (or several) pair(s) of Higgs doublets and
adjoint particles can get GUT/string-scale masses
via high-dimensional operators, we obtain
the Minimal Supersymmetric Standard Model (MSSM) in
the observable sector
after we decouple the heavy particles around the GUT/string scale.
Thus, choosing the suitable grand unified gauge coupling
by adjusting the string scale,
 we can explain the observed low energy gauge couplings 
via renormalization group equation (RGE) running. However, 
how to generate the up-type quark Yukawa couplings, which are
forbidden by the global $U(1)$ symmetry,
 deserves further study.

Furthermore, we consider the flipped $SU(5)$ models.
In order to have at least one pair of Higgs
fields ${\bf 10}$ and ${\bf \overline{10}}$, we must have
the symmetric representations, and then the net number
of ${\bf \overline{5}}$ and ${\bf 5}$ can not be three
if the net number of ${\bf 10}$ and ${\bf \overline{10}}$ is three
due to the non-abelian anomaly free condition.
For the first time, we construct the flipped $SU(5)$ model 
with exact three ${\bf 10}$, and the flipped $SU(5)$ model in which all 
the Yukawa couplings are allowed by the global $U(1)$
symmetries. We will also comment on two more
flipped $SU(5)$ models, and try to avoid as much extra matter 
as possible.

This paper is organized as follows. In Section II, we  briefly
review the intersecting D6-brane model building on Type IIA
orientifolds with flux compactifications. We study the general
conditions for three-family $SU(5)$ model building in Section III.
And we discuss the $SU(5)$ and flipped $SU(5)$ models in Sections IV
and V, respectively.
Discussion and conclusions are given in Section VI.

\section{Flux Model Building on Type IIA orientifolds}

We briefly review the rules for the intersecting D6-brane model
building in Type IIA theory on $\mathbf{T^6/(\Z_2\times \Z_2)}$
orientifold with flux
compactifications~\cite{Villadoro:2005cu,Camara:2005dc}. Because
the model building rules in Type IIA theory on $\mathbf{T^6}$
orientifold with flux compactifications are quite similar, we only
explain the differences for simplicity.

\subsection{Type IIA Theory on $\mathbf{T^6/(\Z_2\times \Z_2)}$ Orientifold}

We consider $\mathbf{T^6}$ to be a six-torus factorized as
$\mathbf{T^6} = \mathbf{T^2} \times \mathbf{T^2} \times
\mathbf{T^2}$ whose complex coordinates are $z_i$, $i=1,\; 2,\; 3$
for the $i$-th two-torus, respectively. The $\theta$ and $\omega$
generators for the orbifold group $\Z_{2} \times \Z_{2}$ act on
the complex coordinates as following
\begin{eqnarray}
& \theta: & (z_1,z_2,z_3) \to (-z_1,-z_2,z_3)~,~ \nonumber \\
& \omega: & (z_1,z_2,z_3) \to (z_1,-z_2,-z_3)~.~\, \label{Z2Z2}
\end{eqnarray}

We implement an orientifold projection $\Omega R$, where $\Omega$
is the world-sheet parity, and $R$ acts on the complex coordinates
as
\begin{equation}
R:(z_1,z_2,z_3)\rightarrow(\overline{z}_1,\overline{z}_2,\overline{z}_3)~.~\,
\end{equation}

Thus, we have four kinds of orientifold 6-planes (O6-planes) under
the actions of $\Omega R$, $\Omega R\theta$, $\Omega R \omega$,
and $\Omega R\theta\omega$, respectively. In addition, we
introduce some stacks of D6-branes which wrap on the factorized
three-cycles. There are two kinds of complex structures consistent
with orientifold projection for a two-torus -- rectangular and
tilted~\cite{CSU,LUII}. If we denote the homology classes of the
three cycles wrapped by $a$ stack of $N_a$ D6-branes as
$n_a^i[a_i]+m_a^i[b_i]$ and $n_a^i[a'_i]+m_a^i[b_i]$ with
$[a_i']=[a_i]+\frac{1}{2}[b_i]$ for the rectangular and tilted
two-tori respectively, we can label a generic one cycle by
$(n_a^i,l_a^i)$ in which $l_{a}^{i}\equiv m_{a}^{i}$ for a
rectangular two-torus while $l_{a}^{i}\equiv
2\tilde{m}_{a}^{i}=2m_{a}^{i}+n_{a}^{i}$ for a tilted
two-torus~\cite{Cvetic:2002pj}. For $a$ stack of $N_a$ D6-branes
along the cycle $(n_a^i,l_a^i)$, we also need to include their
$\Omega R$ images $N_{a'}$ with wrapping numbers $(n_a^i,-l_a^i)$.
For the D6-branes on the top of O6-planes, we count them and their
$\Omega R$ images independently. So, the homology three-cycles for
$a$ stack of $N_a$ D6-branes and its orientifold image $a'$ are
\begin{eqnarray}
[\Pi_a]=\prod_{i=1}^{3}\left(n_{a}^{i}[a_i]+2^{-\beta_i}l_{a}^{i}[b_i]\right),\;\;\;
\left[\Pi_{a'}\right]=\prod_{i=1}^{3}
\left(n_{a}^{i}[a_i]-2^{-\beta_i}l_{a}^{i}[b_i]\right)~,~\,
\end{eqnarray}
where $\beta_i=0$ if the $i$-th two-torus is rectangular and
$\beta_i=1$ if it is tilted. The homology three-cycles wrapped
by the four O6-planes are
\begin{eqnarray}
\Omega R: [\Pi_{\Omega R}]= 2^3 [a_1]\times[a_2]\times[a_3]~,~\,
\end{eqnarray}
\begin{eqnarray}
\Omega R\omega: [\Pi_{\Omega
R\omega}]=-2^{3-\beta_2-\beta_3}[a_1]\times[b_2]\times[b_3]~,~\,
\end{eqnarray}
\begin{eqnarray}
\Omega R\theta\omega: [\Pi_{\Omega
R\theta\omega}]=-2^{3-\beta_1-\beta_3}[b_1]\times[a_2]\times[b_3]~,~\,
\end{eqnarray}
\begin{eqnarray}
\Omega R\theta:  [\Pi_{\Omega
R}]=-2^{3-\beta_1-\beta_2}[b_1]\times[b_2]\times[a_3]~.~\,
\label{orienticycles}
\end{eqnarray}

Therefore, the intersection numbers are
\begin{eqnarray}
I_{ab}=[\Pi_a][\Pi_b]=2^{-k}\prod_{i=1}^3(n_a^il_b^i-n_b^il_a^i)~,~\,
\end{eqnarray}
\begin{eqnarray}
I_{ab'}=[\Pi_a]\left[\Pi_{b'}\right]=-2^{-k}\prod_{i=1}^3(n_{a}^il_b^i+n_b^il_a^i)~,~\,
\end{eqnarray}
\begin{eqnarray}
I_{aa'}=[\Pi_a]\left[\Pi_{a'}\right]=-2^{3-k}\prod_{i=1}^3(n_a^il_a^i)~,~\,
\end{eqnarray}
\begin{eqnarray}
{I_{aO6}=[\Pi_a][\Pi_{O6}]=2^{3-k}(-l_a^1l_a^2l_a^3
+l_a^1n_a^2n_a^3+n_a^1l_a^2n_a^3+n_a^1n_a^2l_a^3)}~,~\,
\label{intersections}
\end{eqnarray}
where $[\Pi_{O6}]=[\Pi_{\Omega R}]+[\Pi_{\Omega
R\omega}]+[\Pi_{\Omega R\theta\omega}]+[\Pi_{\Omega R\theta}]$ is
the sum of O6-plane homology three-cycles wrapped by the four
O6-planes, and $k=\beta_1+\beta_2+\beta_3$ is the total number of
tilted two-tori.

\begin{table}[h]
\renewcommand{\arraystretch}{1.5}
\center
\begin{tabular}{|c||c|}
\hline

Sector & Representation   \\ \hline \hline

$aa$ & $U(N_a /2)$ vector multiplet and 3 adjoint chiral multiplets \\
\hline

$ab+ba$ & $I_{ab}$ $ (\frac{N_a}{2}, \frac{\overline{N_b}}{2})$
chiral multiplets  \\ \hline

$ab'+b'a$ & $I_{ab'}$ $ (\frac{N_a}{2}, \frac{N_b}{2})$ chiral
multiplets  \\ \hline

$aa'+a'a$ & $  \frac 12 (I_{aa'} + \frac 12 I_{aO6}) $
anti-symmetric chiral multiplets  \\

          & $\frac 12 (I_{aa'} - \frac 12 I_{aO6})$
symmetric chiral multiplets  \\ \hline \hline

\end{tabular}
\caption{The general spectrum for the intersecting D6-brane model
building in Type IIA  theory on $\mathbf{T^6/(\Z_2\times \Z_2)}$
orientifold with flux compactifications.}
\label{Spectrum}
\end{table}

For $a$ stack of $N_a$ D6-branes and its $\Omega R$ image, we have
$U(N_a/2)$ gauge symmetry, while for $a$ stack of $N_a$ D6-branes
and its $\Omega R$ image on the top of O6-plane, we obtain
$USp(N_a)$ gauge symmetry. The general spectrum of D6-branes'
intersecting at generic angles, which is valid for both
rectangular and tilted two-tori, is given in Table \ref{Spectrum}.
The four-dimensional $N=1$ supersymmetric models on Type IIA
orientifolds with intersecting D6-branes are mainly constrained in
two aspects: four-dimensional $N=1$  supersymmetry conditions, and
RR tadpole cancellation conditions.

To simplify the notation, we define the following products of
wrapping numbers
\begin{eqnarray}
\begin{array}{rrrr}
A_a \equiv -n_a^1n_a^2n_a^3, & B_a \equiv n_a^1l_a^2l_a^3,
& C_a \equiv l_a^1n_a^2l_a^3, & D_a \equiv l_a^1l_a^2n_a^3, \\
\tilde{A}_a \equiv -l_a^1l_a^2l_a^3, & \tilde{B}_a \equiv
l_a^1n_a^2n_a^3, & \tilde{C}_a \equiv n_a^1l_a^2n_a^3, &
\tilde{D}_a \equiv n_a^1n_a^2l_a^3.\,
\end{array}
\label{variables}
\end{eqnarray}

(1) {\it Four-Dimensional $N=1$  Supersymmetry Conditions}

The four-dimensional $N=1$ supersymmetry  can be preserved by the
orientation projection ($\Omega R$) if and only if the rotation
angle of any D6-brane with respect to any O6-plane is an element
of $SU(3)$~\cite{bdl}, {\it i.~e.}, $\theta_1+\theta_2+\theta_3=0
$ mod $2\pi$, where $\theta_i$ is the angle between the $D6$-brane
and the O6-plane on the $i$-th two-torus. Then the supersymmetry
conditions can be rewritten as~\cite{Cvetic:2002pj}
\begin{eqnarray}
x_A\tilde{A}_a+x_B\tilde{B}_a+x_C\tilde{C}_a+x_D\tilde{D}_a~=~0~,~\,
\label{susyconditions}
\end{eqnarray}
\begin{eqnarray}
 A_a/x_A+B_a/x_B+C_a/x_C+D_a/x_D~<~0~,~\,
\label{susyconditionsII}
\end{eqnarray}
where $x_A=\lambda,\; x_B=\lambda
2^{\beta_2+\beta3}/\chi_2\chi_3,\; x_C=\lambda
2^{\beta_1+\beta3}/\chi_1\chi_3,$ and $x_D=\lambda
2^{\beta_1+\beta2}/\chi_1\chi_2$ in which $\chi_i=R^2_i/R^1_i$ are
the complex structure parameters and $\lambda$ is a positive real
number.

(2) {\it RR Tadpole Cancellation Conditions}

The total RR charges from the D6-branes and O6-planes and from the
metric, NSNS, and RR  fluxes must vanish since the RR field flux
lines are conserved. With the filler branes on the top of the four
O6-planes, we obtain the RR tadpole cancellation
conditions~\cite{Villadoro:2005cu,Camara:2005dc}:
\begin{eqnarray}
2^k N^{(1)} - \sum_a N_a A_a + {1\over 2}(m h_0  + q_1 a_1 + q_2
a_2 + q_3 a_3)& = & 16 ~,~\,
\label{tadodhz2z2I}
\end{eqnarray}
\begin{eqnarray}
-2^{\beta_1} N^{(2)} + \sum_a 2^{-\beta_2-\beta3} N_a B_a +
{1\over 2} (m h_1 - q_1 b_{11} - q_2 b_{21} - q_3 b_{31}) & = &
-2^{4-\beta_2-\beta_3} ~,~\,
\label{tadodhz2z2II}
\end{eqnarray}
\begin{eqnarray}
-2^{\beta_2} N^{(3)} +\sum_a 2^{-\beta_1-\beta_3} N_a C_a +
{1\over 2} ( m h_2 - q_1 b_{12} - q_2 b_{22} - q_3 b_{32}) & = & -
2^{4-\beta_1-\beta_3} ~,~\,
\label{tadodhz2z2III}
\end{eqnarray}
\begin{eqnarray}
-2^{\beta_3} N^{(4)} +\sum_a 2^{-\beta_1-\beta_2} N_a D_a +
{1\over 2} (m h_3 - q_1 b_{13} - q_2 b_{23} - q_3 b_{33}) & = & -
2^{4-\beta_1-\beta_2} ~,~\,
\label{tadodhz2z2IV}
\end{eqnarray}
where $2 N^{(i)}$ are the number of filler branes wrapping along
the $i$-th O6-plane which is defined in Table \ref{orientifold}.
In addition, $a_i$ and $b_{ij}$ arise from the metric fluxes,
$h_0$ and $h_i$ arise from the NSNS fluxes, and $m$ and $q_i$
arise from the RR fluxes. We consider these fluxes ($a_i$,
$b_{ij}$, $h_0$, $h_i$, $m$ and $q_i$) quantized in units of 8 so
that we can avoid the problems with flux Dirac quantization
conditions.

\renewcommand{\arraystretch}{1.4}

\begin{table}[h]
\caption{Wrapping numbers of the four O6-planes.} \vspace{0.4cm}
\begin{center}
\begin{tabular}{|c|c|c|}
\hline
Orientifold Action & O6-Plane & $(n^1,l^1)\times (n^2,l^2)\times (n^3,l^3)$\\
\hline

$\Omega R$& 1 & $(2^{\beta_1},0)\times (2^{\beta_2},0)\times (2^{\beta_3},0)$ \\
\hline

$\Omega R\omega$& 2& $(2^{\beta_1},0)\times (0,-2^{\beta_2})\times (0,2^{\beta_3})$ \\
\hline

$\Omega R\theta\omega$& 3 & $(0,-2^{\beta_1})\times (2^{\beta_2},0)\times (0,2^{\beta_3})$ \\
\hline

$\Omega R\theta$& 4 & $(0,-2^{\beta_1})\times (0,2^{\beta_2})\times (2^{\beta_3},0)$ \\
\hline

\end{tabular}
\end{center}
\label{orientifold}
\end{table}

In this paper, we concentrate on the supersymmetric AdS vacua with
metric, NSNS and RR fluxes~\cite{Camara:2005dc}. For simplicity,
we assume that the K\"ahler moduli $T_i$ satisfy $T_1=T_2=T_3$,
then we obtain $q_1=q_2=q_3\equiv q$ from the superpotential
in~\cite{Camara:2005dc}. To satisfy the Jacobi identities for
metric fluxes, we consider the solution $a_i=a$, $b_{ii}=-b_i$,
and $b_{ji}=b_i$ in which $j\not= i$~\cite{Camara:2005dc}.

To have supersymmetric minima~\cite{Camara:2005dc}, we obtain that
\begin{eqnarray}
3 a {\rm Re}S~=~ b_i {\rm Re} U_i~, ~~{\rm for}~~ i=1, ~2, ~3~,~\,
\label{SUSY-min}
\end{eqnarray}
where
\begin{eqnarray}
{\rm Re}S~\equiv~{{e^{-\phi}}\over {\sqrt {\chi_1 \chi_2
\chi_3}}}~,~ {\rm Re}U_i~\equiv~e^{-\phi} {\sqrt {{\chi_j
\chi_k}\over {\chi_i}}}~,~\,
\end{eqnarray}
where $S$ and $U_i$ are respectively dilaton and complex structure
moduli, $\phi$ is the four-dimensional T-duality invariant
dilaton, and $i \not= j \not= k \not= i$. And then we have
\begin{eqnarray}
b_1={{3a}\over {\chi_2 \chi_3}}~,~ b_2={{3a}\over {\chi_1
\chi_3}}~,~ b_3={{3a}\over {\chi_1 \chi_2}}~.~
\end{eqnarray}

Moreover, there are  consistency conditions
\begin{eqnarray}
3h_i a + h_0 b_i = 0~,~ ~{\rm for}~~ i=1,~2,~3~.~\,
\label{CONS-EQ}
\end{eqnarray}

So we have
\begin{eqnarray}
h_1=-{{h_0}\over {\chi_2 \chi_3}}~,~
h_2=-{{h_0}\over {\chi_1
\chi_3}}~,~
h_3=-{{h_0}\over {\chi_1 \chi_2}}~.~
\end{eqnarray}

Thus, the RR tadpole cancellation conditions can be rewritten as
following
\begin{eqnarray}
2^k N^{(1)} - \sum_a N_a A_a + {1\over 2}(h_0 m + 3 a q)& = & 16
~,~\,
\label{N-tadodhz2z2I}
\end{eqnarray}
\begin{eqnarray}
-2^{\beta_1} N^{(2)} + \sum_a 2^{-\beta_2-\beta3} N_a B_a -
{1\over {2 \chi_2 \chi_3}} (h_0 m + 3 a q) & = &
-2^{4-\beta_2-\beta_3} ~,~\,
\label{N-tadodhz2z2II}
\end{eqnarray}
\begin{eqnarray}
-2^{\beta_2} N^{(3)} +\sum_a 2^{-\beta_1-\beta_3} N_a C_a -
{1\over {2 \chi_1 \chi_3}} (h_0 m + 3 a q)  & = & -
2^{4-\beta_1-\beta_3} ~,~\,
\label{N-tadodhz2z2III}
\end{eqnarray}
\begin{eqnarray}
-2^{\beta_3} N^{(4)} +\sum_a 2^{-\beta_1-\beta_2} N_a D_a -
{1\over {2 \chi_1 \chi_2}} (h_0 m + 3 a q) & = & -
2^{4-\beta_1-\beta_2} ~.~\,
\label{N-tadodhz2z2IV}
\end{eqnarray}

Therefore, if $(h_0 m + 3 a q) < 0$, the supergravity fluxes
contribute negative D6-brane charges to all the RR tadpole
cancellation conditions, and then, the RR tadpole cancellation
conditions give no constraints on the consistent model building
because we can always introduce suitable supergravity fluxes and
some stacks of D6-branes in the hidden sector to cancel the RR
tadpoles. Also, if $(h_0 m + 3 a q) = 0$, the supergravity fluxes
do not contribute to any D6-brane charges, and then do not affect
the RR tadpole cancellation conditions.

In addition, the Freed-Witten anomaly cancellation condition
is~\cite{Camara:2005dc}
\begin{eqnarray}
-2^{-k} h_0 \tilde{A}_a + 2^{-\beta_1} h_1 \tilde{B}_a +
2^{-\beta_2} h_2 \tilde{C}_a + 2^{-\beta_3} h_3 \tilde{D}_a
=0~.~\,
\end{eqnarray}

It can be shown that if Eqs. (\ref{susyconditions}),
(\ref{SUSY-min}), and (\ref{CONS-EQ}) are satisfied, the
Freed-Witten anomaly is automatically cancelled. So, we will not
consider the Freed-Witten anomaly in our model building.

Furthermore, in addition to the above RR tadpole cancellation
conditions, the discrete D-brane RR charges classified by
$\mathbf{\Z_2}$ K-theory groups in the presence of orientifolds,
which are subtle and invisible by the ordinary
homology~\cite{MS,Witten9810188}, should also be taken into
account~\cite{CU}. The K-theory conditions for a
$\mathbf{\Z_2\times \Z_2}$ orientifold are
\begin{equation}
\sum_a 2^{-k} \tilde{A}_a = \sum_a 2^{-\beta_1} \tilde{B}_a =
\sum_a 2^{-\beta_2} \tilde{C}_a  = \sum_a 2^{-\beta_3} \tilde{D}_a
= 0 \textrm{ mod }4 \label{K-charges}~.~\,
\end{equation}

\subsection{Type IIA Theory on $\mathbf{T^6}$ Orientifold}

The intersecting D6-brane model building in Type IIA theory on
$\mathbf{T^6}$ orientifold with flux compactifications is similar
to that on $\mathbf{T^6/(\Z_2\times \Z_2)}$ orientifold. For the
model building rules in the previous subsection, we only need to
make the following changes:

(1) For $a$ stack of $N_a$ D6-branes and its $\Omega R$ image, we
have $U(N_a)$ gauge symmetry, while for $a$ stack of $N_a$
D6-branes and its $\Omega R$ image on the top of O6-plane, we
obtain $USp(2 N_a)$ gauge symmetry. Also, we present the  general
spectrum of D6-branes' intersecting at generic angles in Type IIA
theory on $\mathbf{T^6}$  orientifold in Table \ref{Spectrum-T6}.

(2) We only have the $\Omega R$ O6-planes, so,
$[\Pi_{O6}]=[\Pi_{\Omega R}]$ in Eq. (\ref{intersections}), and
the right-hand sides of Eqs. (\ref{tadodhz2z2II}),
(\ref{tadodhz2z2III}) and (\ref{tadodhz2z2IV}) are zero.

(3) The metric, NSNS and RR fluxes ($a_i$, $b_{ij}$, $h_0$, $h_i$,
$m$ and $q_i$) are quantized in units of 2.

(4) To have three families of the SM fermions, we obtain that at
least one of the three two-tori is tilted. Thus, the right-hand
side of Eq. (\ref{K-charges}) is $0 \textrm{ mod }2$ for the
K-theory conditions in our model
building~\cite{MarchesanoBuznego:2003hp}.

\begin{table}[h]
\renewcommand{\arraystretch}{1.5}
\center
\begin{tabular}{|c||c|}
\hline

Sector & Representation   \\ \hline \hline

$aa$ & $U(N_a)$ vector multiplet and 3 adjoint chiral multiplets \\
\hline

$ab+ba$ & $I_{ab}$ $ (N_a, \overline{N_b})$ chiral multiplets  \\
\hline

$ab'+b'a$ & $I_{ab'}$ $ (N_a, N_b)$ chiral multiplets  \\ \hline

$aa'+a'a$ & $  \frac 12 (I_{aa'} + I_{aO6}) $
anti-symmetric chiral multiplets  \\

  & $\frac 12 (I_{aa'} - I_{aO6})$
symmetric chiral multiplets  \\ \hline

\hline
\end{tabular}

\caption{The general spectrum for the intersecting D6-brane model
building in Type IIA theory on $\mathbf{T^6}$ orientifold with
flux compactifications, in particular,
$I_{aO6}=[\Pi_a][\Pi_{O6}]=-2^{3-k}l_a^1l_a^2l_a^3$.}
\label{Spectrum-T6}
\end{table}

\section{General Conditions for Three-Family $SU(5)$ Model Building}

We would like to construct three-family $SU(5)$ models. First,
we consider the  Type IIA $\mathbf{T^6/(\Z_2\times \Z_2)}$
orientifold. Let us denote  $SU(5)$ stack of D6-branes as $a$
stack. The numbers of the anti-symmetric representations $n_{\bf
10}$ and symmetric representations $n_{\bf 15}$ are
\begin{eqnarray}
n_{\bf 10} &=& 2^{1-k} \left[ (1-2 A_a) \tilde{A}_a + \tilde{B}_a
+\tilde{C}_a + \tilde{D}_a \right]~,~\,
\end{eqnarray}
\begin{eqnarray}
n_{\bf 15} &=& - 2^{1-k} \left[ (1+2 A_a) \tilde{A}_a +
\tilde{B}_a +\tilde{C}_a + \tilde{D}_a \right]~.~\,
\end{eqnarray}

Because we require $n_{\bf 15}=0$, we have
\begin{eqnarray}
\tilde{B}_a +\tilde{C}_a + \tilde{D}_a &=& - (1+2 A_a)
\tilde{A}_a~.~\,
\end{eqnarray}

Then we obtain
\begin{eqnarray}
n_{\bf 10} &=& -2^{3-k} A_a \tilde{A}_a ~.~\,
\end{eqnarray}

Therefore, $k=3$, {\it i.~e.}, all three two-tori must be tilted.
There are four possibilities for $A_a$ and $\tilde{A}_a$:

(1) $A_a=1$ and $\tilde{A}_a=-3$

In this case, we have
\begin{eqnarray}
\tilde{B}_a +\tilde{C}_a + \tilde{D}_a &=& 9~.~\,
\end{eqnarray}

It is easy to show that there is no solution.

(2) $A_a=-1$ and $\tilde{A}_a=3$

In this case, we have
\begin{eqnarray}
\tilde{B}_a +\tilde{C}_a + \tilde{D}_a &=& 3~.~\,
\end{eqnarray}

Up to T-duality and permutations of three different two-tori,
there is one and only one possibility for the wrapping numbers for
$SU(5)$ stacks of D6-branes
\begin{eqnarray}
(1, 3) \times (1, 1) \times (1, -1)~.~\,
\label{SU5-Z2Z2}
\end{eqnarray}

(3) $A_a=3$ and $\tilde{A}_a=-1$

In this case, we have
\begin{eqnarray}
\tilde{B}_a +\tilde{C}_a + \tilde{D}_a &=& 7~.~\,
\end{eqnarray}

Up to T-duality and permutations of three different two-tori,
there is one and only one  possibility for the wrapping numbers
for $SU(5)$ stacks of D6-branes
\begin{eqnarray}
(-3, 1) \times (-1, 1) \times (-1, 1)~.~\,
\end{eqnarray}

(4) $A_a=-3$ and $\tilde{A}_a=1$

In this case, we have
\begin{eqnarray}
\tilde{B}_a +\tilde{C}_a + \tilde{D}_a &=& 5~.~\,
\end{eqnarray}

Up to T-duality and permutations of three different two-tori,
there is one and only one  possibility for the wrapping numbers
for $SU(5)$ stacks of D6-branes
\begin{eqnarray}
(3, -1) \times (1, 1) \times (1, 1)~.~\,
\end{eqnarray}

Second, let us consider Type IIA $\mathbf{T^6}$ orientifold. The
numbers of the anti-symmetric and symmetric
representations are
\begin{eqnarray}
n_{\bf 10} ~=~ 2^{2-k} (1- A_a) \tilde{A}_a ~,~ n_{\bf 15} ~=~ -
2^{2-k} (1+ A_a) \tilde{A}_a ~.~\,
\end{eqnarray}

Because we require  $n_{\bf 15}=0$, we have
\begin{eqnarray}
A_a=-1~.~\,
\end{eqnarray}

Then we obtain
\begin{eqnarray}
n_{\bf 10} &=& 2^{3-k} \tilde{A}_a ~.~\,
\end{eqnarray}

Therefore, we have $k=3$ and $\tilde{A}_a=3$. Up to T-duality and
permutations of three different two-tori, there are four
possibilities for the wrapping numbers for $SU(5)$ stacks of
D6-branes
\begin{eqnarray}
&& (1, -3) \times (1, 1) \times (1, 1)~;~\, \label{WN-SU5} \\
&& (1, 3) \times (1, 1) \times (1, -1)~;~\, \\
&& (1, 3) \times (-1, 1) \times (-1, 1)~;~\, \\
&& (-1, 3) \times (1, 1) \times (-1, 1)~.~\,
\end{eqnarray}

\newpage

\section{$SU(5)$ Models}

In  the previous
$SU(5)$ model building in Type IIA theory on
$\mathbf{T^6/(\Z_2\times \Z_2)}$ orientifold without fluxes, one
can easily show that one can not construct the models with three
anti-symmetric representations and without symmetric
representations~\cite{Cvetic:2002pj,Gmeiner:2006vb}. And then, for the models with
three anti-symmetric representations and some symmetric
representations, the net number of  ${\bf \bar 5}$ and ${\bf 5}$ can not be
 three due to the non-abelian anomaly free
conditions, {\it i.~e.}, one does not have exact three families of
the SM fermions~\cite{Cvetic:2002pj,Gmeiner:2006vb}.

In this Section, we will present $SU(5)$ models with 
three anti-symmetric ${\bf 10}$ representations
and without symmetric ${\bf 15}$ representations.
 Although the net number of ${\bf \bar 5}$ and ${\bf 5}$  is three 
due to the non-Abelian anomaly free condition,
the initial ${\bf \bar 5}$ number is not three.
For a concrete model, we will show that 
after the additional gauge symmetry breaking
via supersymmetry preserving Higgs mechanism, 
the ${\bf \bar 5}$ and ${\bf 5}$ pairs can form the massive vector-like particles with
masses around the GUT/string scale. Then
we will have exact three ${\bf \bar 5}$ and no ${\bf 5}$.
Moreover, we can break the $SU(5)$ gauge symmetry 
 down to the SM gauge symmetry via D6-brane
splitting, and solve the doublet-triplet splitting problem.
If the extra  one pair of Higgs doublets and
adjoint particles can obtain GUT/string-scale masses
via high-dimensional operators, we only have 
 the MSSM in the observable sector below the GUT scale.
And then we can explain the observed low energy gauge
couplings. We also briefly comment on two more
models where the phenomenological discussions are similar.

\subsection{Model SU(5)-I}

We present the D6-brane configurations and intersection numbers for
the Model SU(5)-I in Table \ref{SU5-1}, and its particle spectrum in the
observable and Higgs sectors in Table \ref{SU5-1-spectrum}.
The wrapping numbers for $SU(5)$ stack of D6-branes are equivalent 
to these in Eq.~(\ref{WN-SU5}) by using sign equivalent principle
and interchanging the first and third two-tori~\cite{CLL}.

\begin{table}[h]
\renewcommand{\arraystretch}{1}
\begin{center}
\footnotesize
\begin{tabular}{|@{}c@{}|c||@{}c@{}c@{}c@{}||c|c||c|@{}c@{}|c|@{}c@{}||
@{}c@{}|@{}c@{}|@{}c@{}|@{}c@{}|@{}c@{}|@{}c@{}|@{}c@{}|@{}c@{}|@{}c@{}|}
\hline

stk & $N$ & ($n_1$,$l_1$) & ($n_2$,$l_2$) & ($n_3$,$l_3$) & A & S
& $b$ & $b'$ & $c$ & $c'$ & $d$ & $d'$ & $e$ & $e'$ & $f$ & $f'$ &
$g$ & $g'$ & $O6$
  \\ \hline \hline
$a$ & 5 & ( 1, 1) & (-1,-1) & (-1, 3) & 3 & 0 & 0(2) & -3 & -3 &
0(6) & 3 & 0(0) & 0(3) & 0(1) & 1 & - & -3 & - & 3
\\ \hline

$b$ & 1 & ( 0, 2) & ( 1,-3) & ( 1,-3) & -9 & 9 & - & - & -9 & 0(3)
& -3 & 0(2) & 3 & 0(1) & 2 & - & 0(3) & -  & -18   \\ \hline

$c$ & 1 & ( 1,-1) & ( 1, 3) & ( 2, 0) & 0 & 0 & - & - & - & - &
0(6) & 3 & -3 & 0(3) & -2 & - & 0(1) & - & 0(3)
\\ \hline

$d$ & 1 & (-1, 1) & ( 1,-1) & (-1,-3) & -3 & 0 & - & - & - & - & -
& - & 0(1) & 0(3) & -1 & - & 3 & - & -3   \\ \hline

$e$ & 1 & (-1,-1) & ( 0, 2) & ( 1, 3) & 3 & -3 & - & - & - & - & -
& - & - & - & 0(1) & - & 0(3) & - & 6   \\ \hline

$f$ & 6 & ( 2, 0) & ( 0,-2) & ( 0, 2) & 0 & 0 & - & - & - & - & -
& - & - & - & - & - & - & - & -   \\ \hline

$g$ & 4 & ( 0,-2) & ( 0, 2) & ( 2, 0) & 0 & 0 & - & - & - & - & -
& - & - & - & - & - & - & - & -   \\ \hline

$O6$ & 2 & ( 2, 0) & ( 2, 0) & ( 2, 0) & - & - & - & - & - & - & -
& - & - & - & - & - & - & - & -   \\ \hline
\end{tabular}
\caption{D6-brane configurations and intersection numbers
for the Model SU(5)-I on Type IIA $\mathbf{T^6}$ orientifold.
The complete gauge symmetry is
$U(5) \times U(1)^4 \times USp(12) \times USp(8) \times
USp(4)$, and the complex structure parameters are
$\chi_1 =2\sqrt{3/5}, \;\; \chi_2=2\sqrt{1/15}$, and $\chi_3
=2\sqrt{15}/9$.
To satisfy the RR tadpole cancellation conditions,
we choose $h_0=-4(3q+2)$, $a=8$, and $m=2$. }
\label{SU5-1}
\end{center}
\end{table}

\newpage

\begin{table}
\renewcommand{\arraystretch}{1}
\begin{center}
\footnotesize
\begin{tabular}{|@{}c@{}||@{}c@{}||@{}c@{}|@{}c@{}|@{}c@{}|@{}c@{}|
@{}c@{}||@{}c@{}|@{}c@{}|@{}c@{}|@{}c@{}|@{}c@{}|} \hline

 Rep. & Multi. &$U(1)_a$&$U(1)_b$&$U(1)_c$& $U(1)_d$
& $U(1)_e$ & $U(1)_1$ & $U(1)_2$ & $U(1)_3$ & $U(1)_4$ & $U(1)_{free}$    \\
\hline \hline

$({\bf 10,1})$ & 3 & 2 & 0 & 0 & 0 & 0 & 10 & 10 & -30 & 30 & 2   \\

$({\bf \bar{5}}_a ,{\bf \bar{1}}_b)$ & 3 & -1 & -1 & 0 & 0 & 0 & -7 & -5 & 15 & -15 & -2  \\

$({\bf \bar{1}}_b ,{\bf 1}_c)$ & 3 & 0 & -1 & 1 & 0 & 0 & -4 & 6 & 0 & 18 & 0 \\
\hline \hline

$({\bf 5}_a,{\bf \bar{1}}_b)^\star$ & 2 & 1 & -1 & 0 & 0 & 0 & 3 & 5 & -15 &
33 & 0   \\

$({\bf \bar{5}}_a,{\bf 1}_b)^\star$ & 2 & -1 & 1 & 0 & 0 & 0 & -3 & -5 & 15 &
-33 & 0  \\ \hline \hline

$({\bf \bar{5}}_a,{\bf 1}_c)$ & 3 & -1 & 0 & 1 & 0 & 0 & -7 & 1
& 15 & -15 & 0 \\

$({\bf 5}_a,{\bf \bar{1}}_d)$ & 3 & 1 & 0 & 0 & -1 & 0 & 6 & 6 & -18 & 18 & -4
\\ \hline

$({\bf 5}_a,{\bf 12}_f)$ & 1 & 1 & 0 & 0 & 0 & 0 & 5 & 5 & -15 & 15 & 1
\\

$({\bf \bar{5}}_a,{\bf 8}_g)$ & 3 & -1 & 0 & 0 & 0 & 0 & -5 & -5 & 15 & -15 &
-1 \\

$({\bf 5}_a,{\bf 4}_{O6})$ & 3 & 1 & 0 & 0 & 0 & 0 & 5 & 5 & -15 & 15 & 1
\\ \hline

$({\bf 1}_e,{\bf 4}_{O6})$ & 6 & 0 & 0 & 0 & 0 & 1 & 0 & -2 & 0 & 6 & 3 \\ 

$({\bf \bar{1}}_b,{\bf 1}_c)$ & 6 & 0 & -1 & 1 & 0 & 0 & -4 & 6 & 0 & 18 & 0  \\

$({\bf \bar{1}}_b,{\bf 1}_d)$ & 3 & 0 & -1 & 0 & 1 & 0 & -3 & -1 & 3 & 15 & 4 \\

$({\bf 1}_b,{\bf \bar{1}}_e)$ & 3 & 0 & 1 & 0 & 0 & -1 & 2 & 2 & 0 & -24 & -2  \\

$({\bf 1}_c, {\bf 1}_d)$ & 3 & 0 & 0 & 1 & 1 & 0 & -3 & 5 & 3 & -3 & 6 \\

$({\bf \bar{1}}_c, {\bf 1}_e)$ & 3 & 0 & 0 & -1 & 0 & 1 & 2 & -8 & 0 & 6 & 2 \\

\hline\hline

$({\bf 12}_f,{\bf 8}_{g})^\star$ & 4+4 & 0 & 0 & 0 & 0 & 0 & 0 & 0 & 0 & 0 & 0 \\

$({\bf 8}_g,{\bf 4}_{O6})^\star$ & 4+4 & 0 & 0 & 0 & 0 & 0 & 0 & 0 & 0 & 0 & 0
\\ \hline

$({\bf 1}_c,{\bf \bar{1}}_d)^\star$ & 6 & 0 & 0 & 1 & -1 & 0 & -1 & 7 & -3 & 3 & -4 \\

$({\bf \bar{1}}_c,{\bf 1}_d)^\star$ & 6 & 0 & 0 & -1 & 1 & 0 & 1 & -7 & 3 & -3 & 4 
\\ \hline

$({\bf 1}_e,{\bf 12}_f)^\star$ & 1 & 0 & 0 & 0 & 0 & 1 & 0 & -2 & 0 & 6 & 3 \\

$({\bf \bar{1}}_e,{\bf 12}_f)^\star$ & 1 & 0 & 0 & 0 & 0 & -1 & 0 & 2 & 0 & -6 & -3 
\\ \hline

$({\bf 1}_e,{\bf 8}_g)^\star$ & 1 & 0 & 0 & 0 & 0 & 1 & 0 & -2 & 0 & 6 & 3 \\

$({\bf \bar{1}}_e,{\bf 8}_g)^\star$ & 1 & 0 & 0 & 0 & 0 & -1 & 0 & 2 & 0 & -6 & -3
 \\ \hline \hline

\multicolumn{12}{|c|}{Additional chiral and non-chiral Matter}\\
\hline

\end{tabular}
\caption{The particle spectrum in the
observable and Higgs sectors in the Model SU(5)-I
 with the four global $U(1)$s
from the Green-Schwarz mechanism.  The  $\star'd$ representations
indicate vector-like matter. }
\label{SU5-1-spectrum}
\end{center}
\end{table}

The four global $U(1)$'s from the additional $U(1)$ gauge symmetry
breaking due to
the Green-Schwarz mechanism are
\begin{eqnarray}
U(1)_1&=& 5U(1)_a + 2 U(1)_b - 2U(1)_c - U(1)_d~,~\, \nonumber \\
U(1)_2&=& 5U(1)_a + 6 U(1)_c - U(1)_d - 2 U(1)_e~,~\, \nonumber \\
U(1)_3&=& -15 U(1)_a + 3 U(1)_d~,~\, \nonumber \\
U(1)_4&=& 15 U(1)_a - 18 U(1)_b - 3 U(1)_d + 6U(1)_e~.~\,
\end{eqnarray}
And the anomaly-free $U(1)$ is
\begin{equation}
U(1)_{free}=U(1)_a+U(1)_b+U(1)_c+5U(1)_d+3U(1)_e~.~\,
\end{equation}

In this model, we have three ${\bf 10}$ representations,
thirty ${\bf \bar 5}$ representations, and twenty-seven ${\bf 5}$ representations
for $SU(5)$. Then, the net number of ${\bf \bar 5}$ and ${\bf 5}$ is three.
So, the key question is whether we can give the GUT/string-scale
vector-like masses to twenty-seven pairs of ${\bf \bar 5}$ and ${\bf 5}$.

Let us discuss how to decouple the vector-like particles
via supersymmetry preserving Higgs mechanism. 
We have the following superpotential from three-point functions:
\begin{eqnarray}
W_3 &=& y^A_{ijk} ({\bf \bar{5}}_a,{\bf 1}_c)_i ({\bf 5}_a,{\bf \bar{1}}_d)_j 
({\bf \bar{1}}_c,{\bf 1}_d)_k + 
y^B_{ik} ({\bf \bar{5}}_a,{\bf 8}_g)_i ({\bf 5}_a,{\bf 12}_f) ({\bf 12}_f,{\bf 8}_{g})_k
\nonumber \\
&& +y^C_{ijk} ({\bf \bar{5}}_a,{\bf 8}_g)_i ({\bf 5}_a,{\bf 4}_{O6})_j ({\bf 8}_g,{\bf 4}_{O6})_k~.~\,
\end{eqnarray}
After the Higgs fields $({\bf \bar{1}}_c,{\bf 1}_d)_k$ obtain 
vacuum expectation values (VEVs), we can
give the vector-like masses to three pairs of 
$({\bf \bar{5}}_a,{\bf 1}_c)_i$ and $({\bf 5}_a,{\bf \bar{1}}_d)_j$ because
the $({\bf \bar{5}}_a,{\bf 1}_c)_i$, $({\bf 5}_a,{\bf \bar{1}}_d)_j$ and
three of six $({\bf \bar{1}}_c,{\bf 1}_d)_k$ arises from the
intersections on the third two-torus.
In addition, after the Higgs fields $({\bf 12}_f,{\bf 8}_{g})_k$
and $({\bf 8}_g,{\bf 4}_{O6})_k$ obtain VEVs, we can give 
vector-like masses to eight pairs of ${\bf \bar 5}$ and ${\bf 5}$ in 
$({\bf \bar{5}}_a,{\bf 8}_g)_i$ and $({\bf 5}_a,{\bf 12}_f)$, and
to four pairs of ${\bf \bar 5}$ and ${\bf 5}$ in 
$({\bf \bar{5}}_a,{\bf 8}_g)_i$ and $({\bf 5}_a,{\bf 4}_{O6})_j$, respectively.

To further give vector-like masses to additional
twelve pairs of ${\bf \bar 5}$ and ${\bf 5}$, we introduce the
 following superpotential from four-point functions:
\begin{eqnarray}
W_3 &=& 
y^D_{ik} ({\bf \bar{5}}_a,{\bf 8}_g)_i ({\bf 5}_a,{\bf 12}_f) 
({\bf 1}_e,{\bf 12}_f) ({\bf \bar{1}}_e,{\bf 8}_g)_k
+ y^{D\prime}_{ik} ({\bf \bar{5}}_a,{\bf 8}_g)_i ({\bf 5}_a,{\bf 12}_f) 
({\bf \bar{1}}_e,{\bf 12}_f) ({\bf 1}_e,{\bf 8}_g)_k 
\nonumber \\ &&
+y^E_{ijkl} ({\bf \bar{5}}_a,{\bf 8}_g)_i ({\bf 5}_a,{\bf 4}_{O6})_j 
({\bf \bar{1}}_e,{\bf 8}_g)_k ({\bf 1}_e,{\bf 4}_{O6})_l~.~\,
\end{eqnarray}
We point out that $({\bf \bar{5}}_a,{\bf 8}_g)_i$, $({\bf 5}_a,{\bf 4}_{O6})_j$, 
$({\bf \bar{1}}_e,{\bf 8}_g)_k$, and three of six $({\bf 1}_e,{\bf 4}_{O6})_l$
arise from the intersections on the third two-torus.
If we give VEVs to the Higgs fields
$({\bf 1}_e,{\bf 12}_f)$, $({\bf \bar{1}}_e,{\bf 8}_g)_k$, 
$({\bf \bar{1}}_e,{\bf 12}_f)$,
$({\bf 1}_e,{\bf 8}_g)_k$, and $({\bf 1}_e,{\bf 4}_{O6})_l$, we can generate the
vector-like masses for the rest twelve pairs of ${\bf \bar 5}$ and ${\bf 5}$
in $({\bf \bar{5}}_a,{\bf 8}_g)_i$ and 
$({\bf 5}_a,{\bf 12}_f)$/$({\bf 5}_a,{\bf 4}_{O6})_j$.
Therefore, we only have three ${\bf \bar 5}$ and
do not have ${\bf 5}$ after the Higgs mechanism
at the GUT/string scale. Note that there are three 
($U(1)_e$ symmetric) 
singlets with charge ${\bf -2}$ and six $({\bf 1}_e,{\bf 4}_{O6})_l$
with charge ${\bf +1}$ under the $U(1)_e$ gauge symmetry,
we obtain that the D-flatness for $U(1)_e$ gauge symmetry
can be preserved if we give VEVs to these singlets.
And the D-flatness for other broken gauge symmetries 
can be preserved because all the other
relevant Higgs particles are vector-like. 
Also, it is obvious that we have the F-flatness 
for above superpotential. Thus, the Higgs mechanism
can preserve supersymmetry.

To break the $SU(5)$ gauge symmetry down to the
SM gauge symmetry, we split the $a$ stack of
D6-branes into $a_3$ and $a_2$ stacks with respectively 3 and 2 D6-branes.
To have the vector-like MSSM Higgs doublets, we assume that 
 the $a_2$ and $b$  stacks of D6-branes are parallel and
on the top of each other on the third two-torus.
 Then we obtain two pairs of vector-like Higgs doublets
$({\bf 2}_{a_2}, {\bf {\bar 1}}_b)_j$ and 
$({\bf {\bar 2}}_{a_2}, {\bf 1}_b)_j$ ($j=1, 2$)
with quantum numbers $({\bf { 1}, 2, 1/2})$ and $({\bf { 1},
{\bar 2}, -1/2})$ under $SU(3)_C\times SU(2)_L \times U(1)_Y$
 gauge symmetry. We also assume that
the  $a_3$ and $b$  stacks of D6-branes are not 
on the top of each other on the third two-torus. So,
the vector-like triplets will obtain the masses around
the string scale, and the
doublet-triplet splitting problem is solved.
Therefore, below the GUT scale, we have SM
gauge symmetry, three families of the SM fermions,
two pairs of Higgs doublets, and three adjoint particles
for each gauge symmetry in the observable sector.

Suppose that one pair of the Higgs doublets and adjoint particles 
obtain the GUT/string-scale vector-like masses
via high-dimensional operators, we only have the MSSM below the
GUT scale. And then, if we choose the suitable grand unified gauge coupling
by adjusting the string scale $M_S$,
the observed low energy gauge couplings
can be generated  via RGE running. 
Let us discuss the gauge coupling and the string scale. For 
a generic stack $\sigma$ of D6-branes, its gauge coupling at the 
string scale is~\cite{Li:2003xb}
\begin{eqnarray}
(g_{YM}^{\sigma})^2 = {{{\sqrt {8\pi}} M_s}\over\displaystyle
{M_{Pl}}} {1\over\displaystyle {\prod_{i=1}^3 {\sqrt
{\left(n_{\sigma}^{i} \right)^2\chi_i^{-1} + \left(2^{-\beta_i}
l_{\sigma}^i \right)^2 \chi_i}}}}~.~\, 
\label{coupling}
\end{eqnarray}
So,  the $SU(5)$ gauge coupling $g_{YM}^{a}$ at the string scale is
\begin{eqnarray}
(g_{YM}^{a})^2 ~=~ {{(375\pi^2)^{1/4}}\over\displaystyle 4}
{{M_S}\over {M_{Pl}}} ~\simeq~ {{2M_S}\over {M_{Pl}}}~.~\, 
\label{SU5-coupling}
\end{eqnarray}
Thus, we can have the suitable grand unified gauge coupling $g_{YM}^{a}$
by adjusting the string scale. As an example,
to have the MSSM unified gauge coupling $g_{MSSM}$ which is about $1/{\sqrt 2}$,
we choose $M_S \simeq M_{Pl}/4$ which is close to the string scale
in the heterotic string theory.

\newpage

%%%%%%%%%%%%%%%%%%%%%%%%%%%%%%%%%%%%%%%%%%%%%%%%%%%%%%%%%%%%%%%%%%%%%%%%%
%%%%%%%%%%%%%%%%%%%%%%%%%%%%%%%%%%%%%%%%%%%%%%%%%%%%%%%%%%%%%%%%%%%%%%%%%

\subsubsection{Comment on Other Models}

We present the D6-brane configurations and intersection numbers for
the Model SU(5)-II on Type IIA $\mathbf{T^6}$ orientifold
and the Model SU(5)-III on Type IIA $\mathbf{T^6/(\Z_2\times \Z_2)}$
orientifold in Tables \ref{SU5-2} and \ref{SU5-3}, respectively.
The wrapping numbers for $SU(5)$ stack of D6-branes in the Model SU(5)-II
are the same as these in the Model SU(5)-I, and the 
wrapping numbers for $SU(5)$ stack of D6-branes in the Model SU(5)-III
are given in Eq.~(\ref{SU5-Z2Z2}).
Similar to the Model SU(5)-I, we have three ${\bf 10}$ representations,
and three ${\bf \bar 5}$ representations after the additional
gauge symmetry breaking by the supersymmetry 
preserving Higgs mechanism. 
If the  extra  Higgs doublets and
adjoint particles obtain GUT/string-scale masses
via high-dimensional operators, we can explain the
observed low energy gauge couplings via RGE running.

\begin{table}[h]
\renewcommand{\arraystretch}{1}
\begin{center}
\footnotesize
\begin{tabular}{|@{}c@{}|c||@{}c@{}c@{}c@{}||c|c||c|@{}c@{}|c|@{}c@{}||
c|@{}c@{}|@{}c@{}|@{}c@{}|@{}c@{}|@{}c@{}|c|c|@{}c@{}|@{}c@{}|@{}c@{}|@{}c@{}|@{}c@{}|}
\hline

stk & $N$ & ($n_1$,$l_1$) & ($n_2$,$l_2$) & ($n_3$,$l_3$) & A & S
& $b$ & $b'$ & $c$ & $c'$ & $d$ & $d'$ & $e$ & $e'$ & $f$ & $f'$ &
$g$ & $g'$ & $h$ & $h'$ & $i$ & $i'$ & $O6$
  \\ \hline \hline
$a$ & 5 & ( 1, 1) & (-1,-1) & (-1, 3) & 3 & 0 & -3 & 0(6) & -2 &
-2 & 4 & 0(1) & 5 & 0(2) & 12 & 0(9) & -5 & -8 & 1 & - & -3 & - &
3     \\ \hline

$b$ & 1 & ( 1,-1) & ( 1, 3) & ( 2, 0) & 0 & 0 & - & - & 3 & 0(1) &
-2 & 2 & 0(16) & 2 & -6 & 3 & -2 & 1 & -2 & - & 0(1) & - & 0(3)
\\ \hline

$c$ & 1 & ( 0, 2) & ( 1,-3) & ( 1,-1) & -3 & 3 & - & - & - & - &
-3 & 0(6) & -1 & 8 & -10 & -8 & 2 & -1 & 2 & - & 0(1) & -  & -6   \\
\hline

$d$ & 1 & (-3, 1) & ( 1,-1) & (-1,-1) & -2 & -1 & - & - & - & - &
- & - & 2 & -10 & 0(2) & -5 & 5 & 8 & -2 & - & 3 & - & -1   \\
\hline

$e$ & 1 & ( 1,-1) & ( 3,-7) & ( 0, 2) & -7 & 7 & - & - & - & - & -
& - & - & - & -14 & -35 & 18 & 9 & 0(3) & - & 6 & - & -14   \\
\hline

$f$ & 1 & ( 2, 0) & (-1, 1) & (-7,-3) & 0 & 0 & - & - & - & - & -
& - & - & - & - & - & 3 & 24 & 0(7) & - & 6 & - & 0(3)   \\ \hline

$g$ & 1 & ( 1,-3) & ( 0, 2) & ( 3, 1) & 3 & -3 & - & - & - & - & -
& - & - & - & - & - & - & - & 0(9) & - & 0(1) & - & 6   \\ \hline

$h$ & 4 & ( 2, 0) & ( 0,-2) & ( 0, 2) & 0 & 0 & - & - & - & - & -
& - & - & - & - & - & - & - & - & - & - & - & -   \\ \hline

$i$ & 4 & ( 0,-2) & ( 0, 2) & ( 2, 0) & 0 & 0 & - & - & - & - & -
& - & - & - & - & - & - & - & - & - & - & - & -   \\ \hline

$O6$ & 2 & ( 2, 0) & ( 2, 0) & ( 2, 0) & - & - & - & - & - & - & -
& - & - & - & - & - & - & - & - & - & - & - & -  \\ \hline
\end{tabular}
\caption{D6-brane configurations and intersection numbers
for the Model SU(5)-II on Type IIA $\mathbf{T^6}$ orientifold.
The complete gauge symmetry is 
$U(5) \times U(1)^6 \times USp(8) \times USp(8) \times USp(4)$,
and the complex structure parameters are
$\chi_1 =6/\sqrt{7}, \;\; \chi_2=2/\sqrt{7}$, and $ \chi_3
=2\sqrt{7}/3$.
To satisfy the RR tadpole cancellation conditions,
we choose $h_0=-12(3q+2)$, $a=24$, and $m=2$.}
\label{SU5-2}
\end{center}
\end{table}

\newpage
%%%%%%%%%%%%%%%%%%%%%%%%%%%%%%%%%%%%%%%%%%%%%%%%%%%%%%%%%%%%%%%

\begin{table}[h]
\renewcommand{\arraystretch}{1}
\begin{center}
\footnotesize
\begin{tabular}{|@{}c@{}|c||@{}c@{}c@{}c@{}||c|c||c|@{}c@{}|c|@{}c@{}||
c|c|c|c|@{}c@{}|@{}c@{}|c|c|@{}c@{}|@{}c@{}|@{}c@{}|@{}c@{}|@{}c@{}|@{}c@{}|@{}c@{}|@{}c@{}|@{}c@{}|}
\hline

stk & $N$ & ($n_1$,$l_1$) & ($n_2$,$l_2$) & ($n_3$,$l_3$) & A & S
& $b$ & $b'$ & $c$ & $c'$ & $d$ & $d'$ & $e$ & $e'$ & $f$ & $f'$ &
$g$ & $g'$ & $h$ & $h'$ & $i$ & $i'$ & $j$ & $j'$ & $k$ & $k'$ &
$O6^2$     \\ \hline \hline

$a$ & 10 & ( 1, 3) & ( 1, 1) & ( 1,-1) & 3 & 0 & -3 & 0(2) & -2 &
-2 & 3 & 0 & -40 & -7 & 15 & 0 & 30 & -6 & 5 & -16 & -10 & -10 & 5
& 6 & 3 & 8 & 3
\\ \hline

$b$ & 2 & ( 1,-3) & ( 1, 3) & ( 2, 0) & 0 & 0 & - & - & 3 & 0(1) &
0 & 3 & 0 & 0 & 0 & 27 & -27 & 0 & 0 & 0 & 3 & 24 & 8 & 10 & -8 &
-10 & -6   \\ \hline

$c$ & 2 & ( 0, 2) & ( 1,-3) & ( 3,-1) & 0 & 0 & - & - & - & - & -2
& 2 & -40 & 3 & -24 & 3 & 0 & 6 & -3 & -24 & 0 & 0 & -3 & 6 & 15 &
6 & 6  \\  \hline

$d$ & 2 & ( 1,-3) & ( 1,-1) & ( 1, 1) & -3 & 0 & - & - & - & - & -
& - & 7 & 40 & 0 & -15 & 6 & -30 & 16 & -5 & 10 & 10 & 0 & -5 & -8
& -3 & -3   \\  \hline

$e$ & 2 & (-3,11) & (-3,-11) & ( 2, 0) & 0 & 0 & - & - & - & - & -
& - & - & - & -33 & 330 & 330 & 33 & 0 & 0 & 54 & 243 & 70 & -35 &
-70 & -128 & -66   \\  \hline

$f$ & 2 & ( 1,-3) & (-2, 0) & (-7,-3) & -9 & 9 & - & - & - & - & -
& - & - & - & - & - & 0 & 189 & 168 & -21 & 0 & 147 & 14 & -35 &
-98 & -49 & -42   \\ \hline

$g$ & 2 & ( 2, 0) & ( 1,-3) & ( 7, 3) & -9 & 9 & - & - & - & - & -
& - & - & - & - & - & - & - & -21 & 168 & 0 & -42 & -49 & -98 &
-35 & 24 & 0   \\ \hline

$h$ & 2 & (-3,-7) & (-3, 7) & ( 2, 0) & 0 & 0 & - & - & - & - & -
& - & - & - & - & - & - & - & - & - & 189 & 0 & -56 & -70 & 56 &
70 & 42   \\ \hline

$i$ & 2 & ( 0, 2) & (-3,-7) & ( 7, 3) & 0 & 0 & - & - & - & - & -
& - & - & - & - & - & - & - & - & - & - & - & -35 & 24 & -49 & -98
& -42    \\ \hline

$j$ & 2 & (-1, 7) & (-1, 1) & ( 0, 2) & -3 & 3 & - & - & - & - & -
& - & - & - & - & - & - & - & - & - & - & - & - & - & 0 & 0 & 0
\\ \hline

$k$ & 2 & ( 1, 1) & ( 1, 7) & ( 0,-2) & -3 & 3 & - & - & - & - & -
& - & - & - & - & - & - & - & - & - & - & - & - & - & - & - & 0
\\ \hline

$O6^2$ & 8 & ( 2, 0) & ( 0,-2) & ( 0, 2) & - & - & - & - & - & - &
-
& - & - & - & - & - & - & - & - & - & - & - & - & - & - & - & -  \\
\hline
\end{tabular}
\caption{D6-brane configurations and intersection numbers
for the Model SU(5)-III on Type IIA $\mathbf{T^6/(\Z_2\times \Z_2)}$
 orientifold. The complete gauge symmetry is 
$U(5) \times U(1)^{10} \times USp(8)$,
and the complex structure parameters are
$\chi_1 =2/\sqrt{7}, \;\; \chi_2=2/\sqrt{7}$, and $ \chi_3
=2\sqrt{7}$.
To satisfy the RR tadpole cancellation conditions,
we choose $h_0=-4(3q+8)$, $a=32$, and $m=8$.}
\label{SU5-3}
\end{center}
\end{table}

\newpage

\section{Flipped $SU(5)$ Models}

In the previous flipped $SU(5)\times U(1)_X$ model
building~\cite{Chen:2005ab,Chen:2005mj,Chen:2005cf}, 
the net number of ${\bf 10}$ and ${\bf \overline{10}}$
 and the net number of ${\bf \overline{5}}$ and ${\bf 5}$ are
not three, and at least some Yukawa couplings
(for example, the down-type quark Yukawa couplings)
are forbidden by the global $U(1)$ symmetries.
 So, we would like to construct the better
model. In order to obtain at least one pair of Higgs
fields ${\bf 10}$ and ${\bf \overline{10}}$, we must have
the symmetric representation, and then the net number
of ${\bf \overline{5}}$ and ${\bf 5}$ can not be three
if the net number of ${\bf 10}$ and ${\bf \overline{10}}$ is three.
For the first time, we will present the 
flipped $SU(5)$ model with exact 
three ${\bf 10}$, and the model in which all the
Yukawa couplings in the superpotential
 are allowed by the global $U(1)$ symmetries.
 We will also comment on two more
flipped $SU(5)$ models, and try to avoid as much extra matter 
as possible.

\subsection{Basic Flipped $SU(5)$ Phenomenology}

In a flipped $SU(5)\times U(1)_X$ \cite{AEHN, Barr:1981qv,
FSU(5)N} unified model, the electric charge generator $Q$ is only
partially embedded in $SU(5)$, {\it i.e.}, $Q = T_3 -
\frac{1}{5}Y' + \frac{2}{5}\tilde{Y}$, where $Y'$ is the $U(1)$
internal $SU(5)$ and $\tilde{Y}$ is the external $U(1)_X$ factor.
Essentially, this means that the photon is \lq shared\rq \ between
$SU(5)$ and $U(1)_X$. The SM fermions plus the right-handed
neutrino states reside within the representations $\bar{\bf{5}}$,
$\bf{10}$, and $\bf{1}$ of $SU(5)$, which are collectively
equivalent to a spinor $\bf{16}$ of $SO(10)$.  The quark and
lepton assignments are flipped by $u^c_L$ $\leftrightarrow$
$d^c_L$ and $\nu^c_L$ $\leftrightarrow$ $e^c_L$ relative to a
conventional $SU(5)$ GUT embedding:
\begin{equation}
\bar{f}_{\bf{\bar{5},-\frac{3}{2}}}= \left( \begin{array}{c}
              u^c_1 \\ u^c_2 \\ u^c_3 \\ e \\ \nu_e
                    \end{array} \right) _L ; \;\;\;
F_{\bf{10,\frac{1}{2}}}= \left( \left( \begin{array}{c}
              u \\ d \end{array} \right) _L  d^c_L \;\; \nu^c_L
\right)
              ; \;\;\;
l_{\bf{1,\frac{5}{2}}}=e^c_L~.~\,
\end{equation}
In particular this results in  the $\bf{10}$ containing a neutral
component with the same quantum numbers as $\nu^c_L$.  So we can
spontaneously break the GUT gauge symmetry by using a pair of $\bf{10}$ and
$\overline{\bf{10}}$ of superheavy Higgs where the neutral
components receive a large VEV, $\left\langle
\nu^c_H \right\rangle$= $\left\langle \bar{\nu}^c_H
\right\rangle$,
\begin{equation}
H_{\bf{10,\frac{1}{2}}}=\left\{Q_H,\;d^c_H,\;\nu^c_H \right\};
\;\;\;
\bar{H}_{\bf{\overline{10},-\frac{1}{2}}}=\left\{Q_{\bar{H}},\;d^c_{\bar{H}},\;\nu^c_{\bar{H}}
\right\}~.~\,
\end{equation}
The  spontaneous breaking 
of  electroweak gauge symmetry is generated by the Higgs
doublets $H_2 $ and $ \bar{H}_{\bar{2}} $
\begin{equation}
h_{\bf{5,-1}}=\left\{ H_2,H_3 \right\}; \;\;\;
\bar{h}_{\bf{\bar{5},1}}=\left\{
\bar{H}_{\bar{2}},\bar{H}_{\bar{3}} \right\} ~.~\,
\end{equation}
The flipped $SU(5)$ models have two very nice features which
are generally not found in typical unified models: (i) a natural
solution to the doublet ($H_2$)-triplet($H_3$) splitting problem
of the electroweak Higgs pentaplets $h$ and $\bar{h}$ through the
trilinear couplings of the Higgs fields: $H_{\bf{10}} \cdot
H_{\bf{10}} \cdot h_{\bf{5}} \rightarrow \left\langle \nu^c_H
\right\rangle d^c_H H_3$; (ii) an automatic see-saw mechanism
that provide heavy right-handed neutrino mass through the coupling
to singlet fields $\phi$, $F_{\bf{10}} \cdot {\bar
H}_{\overline{\bf{10}}} \cdot \phi \rightarrow \left\langle
\nu^c_{\bar{H}}\right\rangle \nu^c \phi$.

The generic superpotential $W$ for a flipped $SU(5)$ model is
\begin{equation}
\lambda_1 FFh+\lambda_2 F\bar{f}\bar{h}+\lambda_3 \bar{f}l^c h+
\lambda_4 F\bar{H}\phi +\lambda_5 HHh+\lambda_6
\bar{H}\bar{H}\bar{h}+ \cdots\in W  ~,~\,
\label{sp}
\end{equation}
where the first three terms provide masses for the quarks and leptons,
the fourth is responsible for the heavy right-handed neutrino masses,
and the last two terms are responsible for the doublet-triplet
splitting mechanism \cite{AEHN}.

\subsection{Model FSU(5)-I}

We first construct the Model FSU(5)-I on Type IIA
$\mathbf{T^6}$ orientifold which have exact three ${\bf 10}$
representations. The D6-brane configurations and intersection numbers for
the Model FSU(5)-I are given in Table \ref{FSU5-1}, and its
particle spectrum in the observable sector is given in
Table \ref{FSU5-1-spectrum}.

\begin{table}[h]
\renewcommand{\arraystretch}{1}
\begin{center}
\footnotesize
\begin{tabular}{|@{}c@{}|c||@{}c@{}c@{}c@{}||c|c||c|c|c|c|
c|c|c|c|c|c|c|c|c|c|c|c|c|c|c|c|} \hline

stk & $N$ & ($n_1$,$l_1$) & ($n_2$,$l_2$) & ($n_3$,$l_3$) & A & S
& $b$ & $b'$ & $c$ & $c'$ & $d$ & $d'$ & $e$ & $e'$ & $f$ & $f'$ &
$g$ & $g'$ & $h$ & $h'$ & $i$ & $i'$ & $j$ & $j'$ & $k$ & $k'$
\\ \hline \hline

$a$ & 5 & ( 0, 1) & (-1,-1) & ( 1, 3) & 3 & -3 & -3 & 0(1) & 3 & 0
& 0 & 0 & 0 & 3 & 2 & 1 & 9 & 18 & -2 & -1 & -1 & - & 0 & - & 0 &
-
\\ \hline

$b$ & 1 & ( 1,-1) & ( 0, 2) & ( 1, 3) & -6 & 6 & - & - & -6 & 0 &
-1 & 2 & 0 & -6 & -4 & 0 & -18 & -36 & 0 & 2 & 0 & - & 2 & - & 0 &
-  \\ \hline

$c$ & 1 & ( 1, 1) & ( 1,-1) & ( 2, 0) & 0 & 0 & - & - & - & - & 1
& -2 & 0 & 0 & 0 & -2 & 0 & 0 & 4 & 0 & 2 & - & -2 & - & 0 & - \\
\hline

$d$ & 1 & ( 0, 1) & ( 1,-3) & ( 1,-1) & -3 & 3 & - & - & - & - & -
& - & -2 & -1 & 0 & -3 & -15 & -12 & 0 & 3 & 1 & - & 0 & - & 0 & -
\\  \hline

$e$ & 1 & ( 1,-1) & ( 1, 1) & ( 2, 0) & 0 & 0 & - & - & - & - & -
& - & - & - & 2 & 0 & 0 & 0 & 0 & -4 & -2 & - & 2 & - & 0 & -
\\  \hline

$f$ & 1 & ( 1, 1) & ( 2, 0) & ( 1,-1) & 0 & 0 & - & - & - & - & -
& - & - & - & - & - & -2 & 1 & 3 & 0 & 2 & - & 0 & - & -2 & -
\\ \hline

$g$ & 1 & ( 3,-1) & ( 3, 1) & ( 2, 0) & 0 & 0 & - & - & - & - & -
& - & - & - & - & - & - & - & 20 & -32 & -6 & - & 6 & - & 0 & -
\\ \hline

$h$ & 1 & (-1, 1) & (-1, 3) & ( 0, 2) & -6 & 6 & - & - & - & - & -
& - & - & - & - & - & - & - & - & - & 0 & - & 0 & - & 2 & -
  \\ \hline

$i$ & 3 & ( 1, 0) & ( 0,-2) & ( 0, 2) & 0 & 0 & - & - & - & - & -
& - & - & - & - & - & - & - & - & - & - & - & - & - & - & -  \\
\hline

$j$ & 1 & ( 0,-1) & ( 2, 0) & ( 0, 2) & 0 & 0 & - & - & - & - & -
& - & - & - & - & - & - & - & - & - & - & - & - & - & - & -
\\ \hline

$k$ & 2 & ( 0,-1) & ( 0, 2) & ( 2, 0) & 0 & 0 & - & - & - & - & -
& - & - & - & - & - & - & - & - & - & - & - & - & - & - & -
\\ \hline

\end{tabular}

\caption{D6-brane configurations and intersection numbers
for the Model FSU(5)-I on Type IIA $\mathbf{T^6}$ orientifold.
The complete gauge symmetry is
$U(5) \times U(1)^{7} \times USp(6) \times USp(2) \times
USp(4)$, and the complex structure parameters are
$\chi_1 =1/\sqrt{3}, \;\; \chi_2=2/\sqrt{3}$, and $\chi_3
=2/\sqrt{3}$.
To satisfy the RR tadpole cancellation conditions,
we choose $h_0=-12(3q+2)$, $a=16$, and $m=2$.}
\label{FSU5-1}
\end{center}
\end{table}

\begin{table}[h]
\renewcommand{\arraystretch}{1}
\begin{center}
\footnotesize
\begin{tabular}{|@{}c@{}||@{}c@{}||@{}c@{}|@{}c@{}|@{}c@{}|@{}c@{}|@{}c@{}|@{}c@{}|@{}c@{}|@{}c@{}||@{}c@{}||@{}c@{}|
@{}c@{}|@{}c@{}|@{}c@{}||@{}c@{}|@{}c@{}|@{}c@{}|} \hline

 Rep. & Multi. &$U(1)_a$&$U(1)_b$&$U(1)_c$& $U(1)_d$
& $U(1)_e$ & $U(1)_f$ & $U(1)_g$ & $U(1)_h$ & $U(1)_X$ & $U(1)_1$
& $U(1)_2$ & $U(1)_3$ & $U(1)_4$ & $U(1)_U$ & $U(1)_V$ & $U(1)_W$   \\
\hline \hline

$({\bf 10},{\bf 1})$ & 3 & 2 & 0 & 0 & 0 & 0 & 0 & 0 & 0 & 1 & -10 & 0 & 0 & 30
& 10 & 0 & 0  \\

$({\bf \bar{5}}_a ,{\bf 1}_b)$ & 3 & -1 & 1 & 0 & 0 & 0 & 0 & 0 & 0 & -3 & 5 &
2 & 0 & -21 & -30 & 0 & 4  \\

$({\bf 1}_d, {\bf 1}_h)$ & 3 & 0 & 0 & 0 & 1 & 0 & 0 & 0 & 1 & 5 & 1 & 0 & 2 &
-9 & 50 & -1 & -7
\\ \hline \hline

$({\bf 10},{\bf 1})$ & 1 & 2 & 0 & 0 & 0 & 0 & 0 & 0 & 0 & 1 & -10 & 0 & 0 &
30 & 10 & 0 & 0 \\

$({\bf \overline{10}},{\bf 1})$ & 1 & -2 & 0 & 0 & 0 & 0 & 0 & 0 & 0 & -1 &
10 & 0 & 0 & -30 & -10 & 0 & 0  \\
\hline

$({\bf 5}_a,{\bf 1}_b)^\star$ & 1 & 1 & 1 & 0 & 0 & 0 & 0 & 0 & 0 & -2 & -5 &
2 & 0 & 9 & -20 & 0 & 4   \\

$({\bf \bar{5}}_a, {\bf \bar{1}}_b)^\star$ & 1 & -1 & -1 & 0 & 0 & 0 & 0 & 0 &
0 & 2 & 5 & -2 & 0 & -9 & 20 & 0 & -4
\\ \hline

$({\bf 1}_c, {\bf \bar{1}}_h)$ & 4 & 0 & 0 & 1 & 0 & 0 & 0 & 0 & -1 & 0 & 2 &
-2 & 0 & 6 & 0 & 0 & 1 \\  \hline \hline

$({\bf \overline{15}},{\bf 1})$ & 3 & -2 & 0 & 0 & 0 & 0 & 0 & 0 & 0 & -1 & 10
& 0 & 0 & -30 & -10 & 0 & 0   \\

$({\bf 10},{\bf 1})$ & 2 & 2 & 0 & 0 & 0 & 0 & 0 & 0 & 0 & 1 & -10 & 0 & 0 &
30 & 10 & 0 & 0 \\

$({\bf \overline{10}},{\bf 1})$ & 2 & -2 & 0 & 0 & 0 & 0 & 0 & 0 & 0 & -1 & 10
& 0 & 0 & -30 & -10 & 0 & 0   \\
\hline

\multicolumn{18}{|c|}{Additional chiral and non-chiral Matter}\\
\hline

\end{tabular}
\caption{The particle spectrum in the
observable sector in the Model FSU(5)-I, with 
the four global $U(1)$s from
the Green-Schwarz mechanism.  The $\star'd$ representations
indicate vector-like matter. }
\label{FSU5-1-spectrum}
\end{center}
\end{table}

The $U(1)_X$ in flipped $SU(5)\times U(1)_X$ gauge symmetry is
\begin{eqnarray}
U(1)_X &=& \frac{1}{2}(U(1)_a -5U(1)_b +5U(1)_c +5U(1)_d -5U(1)_e
\nonumber \\ && 
+5U(1)_f + 5U(1)_g +5 U(1)_h)~.~\,
\end{eqnarray}
The other massless $U(1)$'s are:
\begin{eqnarray}
U(1)_U &=& 5U(1)_a -25U(1)_b +25U(1)_c +25U(1)_d +107U(1)_e 
\nonumber \\ && +25U(1)_f
-19U(1)_g +25U(1)_h~,~\,
\end{eqnarray}
\begin{eqnarray}
U(1)_V &=& U(1)_c -2 U(1)_d + U(1)_e + U(1)_f + U(1)_h ~,~\,
\end{eqnarray}
\begin{eqnarray}
U(1)_W &=& 4U(1)_b -6 U(1)_d -10 U(1)_e - U(1)_f +2 U(1)_g
-U(1)_h~.~\,
\end{eqnarray}
And the four global $U(1)$'s are
\begin{eqnarray}
U(1)_1&=& -5U(1)_a + 2 U(1)_c +U(1)_d -2U(1)_e +2 U(1)_f - 6U(1)_g~,~\, \nonumber \\
U(1)_2&=&  2U(1)_b -2U(1)_c +2 U(1)_e +6 U(1)_g~,~\, \nonumber \\
U(1)_3&=& -2 U(1)_f + 2 U(1)_h~,~\, \nonumber \\
U(1)_4&=& 15U(1)_a -6 U(1)_b -3U(1)_d -6 U(1)_h~.~\,
\end{eqnarray}

\newpage

\subsection{Model FSU(5)-II}

We construct the Model FSU(5)-II on Type IIA
$\mathbf{T^6}$ orientifold in which unlike
the previous flipped $SU(5)$ model building~\cite{Chen:2005ab,Chen:2005mj,Chen:2005cf}, 
all the Yukawa couplings are allowed by the global $U(1)$ symmetries.
 The D6-brane configurations and intersection numbers for the 
Model FSU(5)-II are given in Tables \ref{FSU5-2} and \ref{FSU5-2C}, and its
particle spectrum in the observable sector is given in
Table \ref{FSU5-2-spectrum}.

\begin{table}[h]
\renewcommand{\arraystretch}{.9}
\begin{center}
\footnotesize
\begin{tabular}{|@{}c@{}|c||@{}c@{}c@{}c@{}||c|c||c|c|c|c|
c|c|c|c|c|c|} \hline

stk & $N$ & ($n_1$,$l_1$) & ($n_2$,$l_2$) & ($n_3$,$l_3$) & A & S
& $b$ & $b'$ & $c$ & $c'$ & $d$ & $d'$ & $e$ & $e'$ & $f$ & $f'$
\\ \hline \hline

$a$ & 5 & ( 0, 1) & (-1,-1) & ( 3, 1) & 2 & -2 & -3 & -6 & 0(1014)
& 0(864) & 0(242) & 0(392) & 0(6) & 0(0) & -3 & 0(3)
\\ \hline

$b$ & 1 & ( 1, 3) & ( 1, 3) & ( 0,-1) & 18 & -18 & - & - & -114 &
111 & -200 & -425 & -6 & 3 & 0(24) & 0(6) \\ \hline

$c$ & 1 & ( 0, 1) & (25,-1) & ( 3,-25) & -50 & 50 & - & - & - & -
& 0(197192) & 0(193442) & 0(864) & (1014) & 36 & 39  \\
\hline

$d$ & 1 & ( 0, 1) & (-3,-25) & (25, 1) & 50 & -50 & - & - & - & -
& - & - & 0(392) & 0(242) & -250 & 275     \\  \hline

$e$ & 1 & ( 0, 1) & ( 1,-1) & ( 3, 1) & -2 & 2 & - & - & - & - & -
& - & - & - & 0 & 3    \\  \hline

$f$ & 1 & ( 1,-9) & ( 1,-1) & ( 0, 1) & -18 & 18 & - & - & - & - &
- & - & - & - & - & -    \\ \hline

$g$ & 1 & ( 1, 0) & ( 3,-1) & ( 3, 1) & 0 & 0 & - & - & - & - & -
& - & - & - & - & -   \\ \hline

$h$ & 1 & ( 1, 0) & ( 3, 1) & ( 3,-1) & 0 & 0 & - & - & - & - & -
& - & - & - & - & -  \\ \hline

$i$ & 1 & ( 1, 1) & ( 1, 9) & ( 0,-1) & 18 & -18 & - & - & - & - &
- & - & - & - & - & -   \\  \hline

$j$ & 1 & ( 1,-1) & ( 1,-9) & ( 0, 1) & -18 & 18 & - & - & - & - &
- & - & - & - & - & -   \\ \hline

$k$ & 1 & ( 1,-1) & (27, 1) & ( 1, 0) & 0 & 0 & - & - & - & - & -
& - & - & - & - & -    \\ \hline

$l$ & 1 & ( 1, 1) & (27,-1) & ( 1, 0) & 0 & 0 & - & - & - & - & -
& - & - & - & - & -    \\ \hline

$O6$ & 8 & ( 1, 0) & ( 2, 0) & ( 1, 0) & - & - & - & - & - & - & -
& - & - & - & - & -   \\ \hline
\end{tabular}
\caption{D6-brane configurations and intersection numbers (Part 1)
for the Model FSU(5)-II on Type IIA $\mathbf{T^6}$ orientifold.
The complete gauge symmetry is
$U(5) \times U(1)^{11} \times USp(16)$,
 and the complex structure parameters are
$\chi_1 =\sqrt{3}/27, \;\; \chi_2=2\sqrt{3}$, and $\chi_3 =\sqrt{3}$.
To satisfy the RR tadpole cancellation conditions,
we choose $h_0=-6(q+2)$, $a=24$, and $m=12$.}
\label{FSU5-2}
\end{center}
\end{table}

\begin{table}[h]
\renewcommand{\arraystretch}{.9}
\begin{center}
\footnotesize
\begin{tabular}{|@{}c@{}|c||@{}c@{}c@{}c@{}||c|c||c|c|c|c|
c|c|c|c|c|c|c|} \hline

stk & $N$ & ($n_1$,$l_1$) & ($n_2$,$l_2$) & ($n_3$,$l_3$) & $g$ &
$g'$ & $h$ & $h'$ & $i$ & $i'$ & $j$ & $j'$ & $k$ & $k'$ & $l$ &
$l'$ & $O6$  \\ \hline \hline

$a$ & 5 & ( 0, 1) & (-1,-1) & ( 3, 1) & 0 & 6 & 6 & 0 & -12 & -15
& -15 & -12 & 13 & 14 & 14 & 13 & 1
\\ \hline

$b$ & 1 & ( 1, 3) & ( 1, 3) & ( 0,-1) & 45 & 36 & 36 & 45 & 0 & 0
& 0 & 0 & 160 & 82 & 82 & 160 & 9
\\ \hline

$c$ & 1 & ( 0, 1) & (25,-1) & ( 3,-25) & 858 & -1008 & -1008 & 858
& 339 & 336 & 0 & 0 & 160 & 82 & 82 & 160 & -25  \\  \hline

$d$ & 1 & ( 0, 1) & (-3,-25) & (25, 1) & -858 & 1008 & 1008 & -858
& -25 & -650 & -650 & -25 & 336 & 339 & 339 & 336 & 25    \\
\hline

$e$ & 1 & ( 0, 1) & ( 1,-1) & ( 3, 1) & -6 & 0 & 0 & -6 & 15 & 12
& 12 & 15 & -14 & -13 & -13 & -14 & -1   \\  \hline

$f$ & 1 & ( 1,-9) & ( 1,-1) & ( 0, 1) & -27 & -54 & -54 & -27 & 0
& 0 & 0 & 0 & -112 & -130 & -130 & -112 & -9   \\ \hline

$g$ & 1 & ( 1, 0) & ( 3,-1) & ( 3, 1) & - & - & 0 & 0 & -42 & 39 &
39 & -42 & 15 & -12 & -12 & 15 & 0  \\ \hline

$h$ & 1 & ( 1, 0) & ( 3, 1) & ( 3,-1) & - & - & - & - & -39 & 42 &
42 & -39 & 12 & -15 & -15 & 12 & 0 \\ \hline

$i$ & 1 & ( 1, 1) & ( 1, 9) & ( 0,-1) & - & - & - & - & - & - & 0
& 0 & 242 & 0 & 0 & 242 & 0  \\  \hline

$j$ & 1 & ( 1,-1) & ( 1,-9) & ( 0, 1) & - & - & - & - & - & - & -
& - & 0 & -242 & -242 & 0 & -9  \\ \hline

$k$ & 1 & ( 1,-1) & (27, 1) & ( 1, 0) & - & - & - & - & - & - & -
& - & - & - & 0 & 0 & 0   \\ \hline

$l$ & 1 & ( 1, 1) & (27,-1) & ( 1, 0) & - & - & - & - & - & - & -
& - & - & - & - & - & 0   \\ \hline

$O6$ & 8 & ( 1, 0) & ( 2, 0) & ( 1, 0) & - & - & - & - & - & - & -
& - & - & - & - & - & -  \\ \hline
\end{tabular}
\caption{D6-brane configurations and intersection numbers (Part 2)
for the Model FSU(5)-II on Type IIA $\mathbf{T^6}$ orientifold.}
\label{FSU5-2C}
\end{center}
\end{table}

\begin{table}[h]
\renewcommand{\arraystretch}{1}
\begin{center}
\footnotesize
\begin{tabular}{|@{}c@{}||@{}c@{}||@{}c@{}||@{}c@{}|@{}c@{}|@{}c@{}|
@{}c@{}||@{}c@{}|@{}c@{}|c|} \hline

Rep. & Multi.  & $U(1)_X$ & $U(1)_1$ & $U(1)_2$ & $U(1)_3$ &
$U(1)_4$ & $U(1)_V$ & $U(1)_W$ & $\; \cdots$   \\
\hline \hline

$({\bf 10},{\bf 1})$ & 3 & 1 & -30 & 0 & 0 & 10 & 0 & 0 & $\cdots$   \\

$({\bf \bar{5}}_a ,{\bf 1}_b)$ & 3 & -3 & 15 & 0 & -1 & 4 & 1 & -36 & $\cdots$  \\

$({\bf \bar{1}}_b, {\bf 1}_d)$ & 3 & 5 & -75 & 0 & 1 & 16 & -1 & 36 & $\cdots$
\\ \hline \hline

$({\bf 10},{\bf 1})$ & 1 & 1 & -30 & 0 & 0 & 10 & 0 & 0 & $\cdots$  \\

$({\bf \overline{10}},{\bf 1})$ & 1 & -1 & 30 & 0 & 0 & -10 & 0 & 0 & $\cdots$  \\
\hline

$({\bf 5}_a,{\bf \bar{1}}_d)^\star$ & 1 & -2 & 60 & 0 & 0 & -20 & 0 & 0 & $\cdots$   \\

$\bar{h}_x$ ($({{\bf \bar{5}}}_a,{\bf 1}_d)^\star$/$({{\bf \bar{5}}}_a,{\bf 1}_f)^\star$)
& 1 & 2 & -60/15 & 0 & 0/1 & 20/-14 & 0/-1 & 0/36 &
$\cdots$ \\ \hline

$({\bf 1}_b ,{\bf 1}_f)$ & 4 & 0 & 0 & 0 & 0 & 0 & 0 & 0 & $\cdots$  \\
\hline \hline

$({\bf \overline{15}},{\bf 1})$ & 2 & -1 & 30 & 0 & 0 & -10 & 0 & 0 & $\cdots$    \\

$({\bf \overline{10}},{\bf 1})$ & 1 & -1 & 30 & 0 & 0 & -10 & 0 & 0 & $\cdots$    \\
\hline

\multicolumn{10}{|c|}{Additional chiral and non-chiral Matter}\\
\hline

\end{tabular}
\caption{The particle spectrum in the
observable sector in the Model FSU(5)-II, with 
the four global $U(1)$s from
the Green-Schwarz mechanism.  The $\star'd$ representations
indicate vector-like matter. }
\label{FSU5-2-spectrum}
\end{center}
\end{table}

The $U(1)_X$ gauge symmetry  is
\begin{eqnarray}
U(1)_X &=& \frac{1}{2}(U(1)_a -5U(1)_b +5U(1)_c +5U(1)_d +5U(1)_e
+5U(1)_f + 5U(1)_g \nonumber \\
&& +5 U(1)_h + 5U(1)_i -5U(1)_j - 5 U(1)_k - 5 U(1)_l)~.~
\end{eqnarray}

The four global $U(1)$'s are:
\begin{eqnarray}
U(1)_1&=& -15U(1)_a + 75 U(1)_c -75 U(1)_d +3U(1)_e -27 U(1)_k + 27U(1)_l~,~\, \nonumber \\
U(1)_2&=& -3U(1)_g + 3U(1)_h + U(1)_k - U(1)_l~,~\, \nonumber \\
U(1)_3&=& -U(1)_b + U(1)_f + 3U(1)_g - 3U(1)_h -U(1)_i + U(1)_j~,~\, \nonumber \\
U(1)_4&=& 5U(1)_a +9 U(1)_b -25U(1)_c + 25 U(1)_d - U(1)_e -
9U(1)_f + 9 U(1)_i - 9 U(1)_j~.~\, \;\;\;\;
\end{eqnarray}

There are seven other massless  $U(1)$'s. As an example, we present two
of them:
\begin{eqnarray}
U(1)_V &=&  U(1)_b - U(1)_f +2 U(1)_g + 2 U(1)_h -2 U(1)_i~,~\, \nonumber \\
U(1)_W &=& -36U(1)_b - 27 U(1)_c +36 U(1)_f +4 U(1)_g+ 29 U(1)_h -3 U(1)_i
+ 75U(1)_l ~.~\,
\end{eqnarray}

This is the first trial flipped $SU(5)$ model
where all the Yukawa couplings in superpotential in Eq. (\ref{sp})
are allowed by the global $U(1)$'s from
the Green-Schwarz mechanism.
  To make the terms like $FFh$ or $HHh$
to be neutral under the global $U(1)$ symmetries,
 we need to set the Higgs pentaplet $h$ from the
intersection between the $N=5$ stack and a stack with large
wrapping numbers (by a factor of 25 due to the flipped $SU(5)$
structure) and therefore we can not avoid extremely large exotic
matter in the spectrum.  In this model the Yukawa terms are:
\begin{eqnarray}
F F h \; &\rightarrow &\;({\bf 10},{\bf 1})
({\bf 10},{\bf 1}) ({\bf 5}_a,{\bf \bar{1}}_d) ~,~\, \nonumber \\
F \bar{f} \bar{h'} \; &\rightarrow &\; ({\bf 10},{\bf 1})
({\bf \bar{5}}_a,{\bf 1}_b)
({\bf \bar{5}}_a,{\bf 1}_f) ~,~\, \nonumber \\
\bar{f} l^c h \; &\rightarrow &\;
({\bf \bar{5}}_a,{\bf 1}_b)
({\bf \bar{1}}_b,{\bf 1}_d)
({\bf 5}_a,{\bf \bar{1}}_d) ~,~\,
\nonumber \\
F \bar{H} \phi \; &\rightarrow &\; ({\bf 10},{\bf 1})
(\overline{\bf {10}},{\bf 1}) ({\bf 1}_b,{\bf 1}_f)~,~\,
\nonumber \\
H H h \; &\rightarrow &\; ({\bf 10},{\bf 1})
({\bf 10},{\bf 1}) ({\bf 5}_a,{\bf \bar{1}}_d) ~,~\, \nonumber \\
\bar{H} \bar{H} \bar{h} \; &\rightarrow &\;
({\bf \overline{{10}}},{\bf 1})({\bf \overline{{10}}},{\bf 1})
({\bf \bar{5}}_a,{\bf 1}_d) ~.~\,
\end{eqnarray}
Because of the structure of Green-Schwarz mechanism in D-brane construction,
to cancel  the global $U(1)$'s charges for all the Yukawa couplings we
expect a mixture state of Higgs pentaplet
$\bar{h}_x=c\bar{h'}+s\bar{h}$ where $\bar{h'}$ is from $F \bar{f}
\bar{h'}$ and $\bar{h}$ is from $\bar{H} \bar{H} \bar{h}$.
However, we may reintroduce the doublet-triplet splitting 
problem.

\subsection{Model FSU(5)-III}

We present the D6-brane configurations and intersection numbers for
the Model FSU(5)-III in Table \ref{FSU5-3}, and its particle spectrum 
in the observable sector in Table \ref{FSU5-3-spectrum}.

\begin{table}[h]
\renewcommand{\arraystretch}{1}
\begin{center}
\footnotesize
\begin{tabular}{|@{}c@{}|c||@{}c@{}c@{}c@{}||c|c||c|@{}c@{}|@{}c@{}|@{}c@{}|
c|c|c|c|c|c|c|c|c|c|c|} \hline

stk & $N$ & ($n_1$,$l_1$) & ($n_2$,$l_2$) & ($n_3$,$l_3$) & A & S
& $b$ & $b'$ & $c$ & $c'$ & $d$ & $d'$ & $e$ & $e'$ & $f$ & $f'$ &
$g$ & $g'$ & $h$ & $h'$ & $O6$
\\ \hline \hline

$a$ & 5 & ( 0, 1) & (-1,-1) & ( 3, 1) & 2 & -2 & -3 & 0(3) & 0(6)
& 0(0) & -3 & -6 & 5 & 4 & 1 & 2 & 0 & 0 & 0 & 0 & 1
\\ \hline

$b$ & 1 & (-1,-3) & ( 1,-1) & ( 0, 1) & -6 & 6 & - & - & 0(3) & 3
& 0 & 0 & -24 & 0 & 12 & -12 & 12 & -9 & -9 & 12 & -3  \\ \hline

$c$ & 1 & ( 0, 1) & ( 1,-1) & ( 3,-1) & -2 & 2 & - & - & - & - & 6
& 3 & -4 & -5 & -2 & -1 & 0 & 0 & 0 & 0 & -1  \\ \hline

$d$ & 1 & ( 1, 1) & ( 1, 3) & ( 0,-1) & 6 & -6 & - & - & - & - & -
& - & -28 & 52 & 40 & -40 & 30 & -33 & -33 & 30 & 3 \\  \hline

$e$ & 1 & ( 1, 3) & ( 9,-1) & ( 1, 0) & 0 & 0 & - & - & - & - & -
& - & - & - & 0 & 0 & -56 & 7 & 7 & -56 & 0  \\  \hline

$f$ & 1 & ( 1,-9) & ( 3, 1) & ( 1, 0) & 0 & 0 & - & - & - & - & -
& - & - & - & - & - & 14 & 35 & 35 & 14 & 0   \\ \hline

$g$ & 1 & ( 0, 1) & ( 7, 1) & (-3,-7) & 14 & -14 & - & - & - & - &
- & - & - & - & - & - & - & - & 0 & 0 & 7  \\ \hline

$h$ & 1 & ( 0, 1) & ( 7,-1) & ( 3,-7) & -14 & 14 & - & - & - & - &
- & - & - & - & - & - & - & - & - & - & -7
\\ \hline

$O6$ & 8 & ( 1, 0) & ( 2, 0) & ( 1, 0) & - & - & - & - & - & - & -
& - & - & - & - & - & - & - & - & - & -   \\ \hline

\end{tabular}
\caption{D6-brane configurations and intersection numbers
for the Model FSU(5)-III on Type IIA $\mathbf{T^6}$ orientifold.
The complete gauge symmetry is
$U(5) \times U(1)^{7} \times USp(16)$,
 and the complex structure parameters are
$\chi_1 =1/9, \;\; \chi_2=6$, and $\chi_3 =1$.
To satisfy the RR tadpole cancellation conditions,
we choose $h_0=-6(3q+2)$, $a=12$, and $m=2$.}
\label{FSU5-3}
\end{center}
\end{table}

\begin{table}[h]
\renewcommand{\arraystretch}{1}
\begin{center}
\footnotesize
\begin{tabular}{|@{}c@{}||@{}c@{}||@{}c@{}|@{}c@{}|@{}c@{}|@{}c@{}|@{}c@{}|@{}c@{}|@{}c@{}|@{}c@{}||@{}c@{}||@{}c@{}|
@{}c@{}|@{}c@{}|@{}c@{}||@{}c@{}|@{}c@{}|@{}c@{}|} \hline

 Rep. & Multi. &$U(1)_a$&$U(1)_b$&$U(1)_c$& $U(1)_d$
& $U(1)_e$ & $U(1)_f$ & $U(1)_g$ & $U(1)_h$ & $U(1)_X$ & $U(1)_1$
& $U(1)_2$ & $U(1)_3$ & $U(1)_4$ & $U(1)_U$ & $U(1)_V$ & $U(1)_W$   \\
\hline \hline

$({\bf 10},{\bf 1})$ & 3 & 2 & 0 & 0 & 0 & 0 & 0 & 0 & 0 & 1 & -30 & 0 & 0 &
10 & 0 & -20 & 70  \\

$({\bf \bar{5}}_a ,{\bf 1}_b)$ & 3 & -1 & 1 & 0 & 0 & 0 & 0 & 0 & 0 & -3 & 15
& 0 & 1 & -8 & 0 & -20 & 70  \\

$({\bf 1}_c, {\bf \bar{1}}_d)$ & 3 & 0 & 0 & 1 & -1 & 0 & 0 & 0 & 0 & 5 & 3 &
0 & 1 & -4 & 0 & -50 & -7
\\ \hline \hline

$({\bf 10},{\bf 1})$ & 1 & 2 & 0 & 0 & 0 & 0 & 0 & 0 & 0 & 1 & -30 & 0 & 0 &
10 & 0 & -20 & 70 \\

$({\bf \overline{10}},{\bf 1})$ & 1 & -2 & 0 & 0 & 0 & 0 & 0 & 0 & 0 & -1 &
30 & 0 & 0 & -10 & 0 & 20 & -70  \\
\hline

$({\bf 5}_a,{\bf \bar{1}}_c)^\star$ & 1 & 1 & 0 & -1 & 0 & 0 & 0 & 0 & 0 & -2
& -18 & 0 & 0 & 6 & 0 & 40 & 42   \\

$({\bf \bar{5}}_a,{\bf 1}_c)^\star$ & 1 & -1 & 0 & 1 & 0 & 0 & 0 & 0 & 0 & 2
& 18 & 0 & 0 & -6 & 0 & -40 & -42
\\ \hline

$({\bf \bar{1}}_c,{\bf 1}_e)$ & 4 & 0 & 0 & -1 & 0 & 1 & 0 & 0 & 0 & 0 & 24 &
-1 & 0 & 1 & 1 & 63 & 7 \\  \hline \hline

$({\bf \overline{15}},{\bf 1})$ & 2 & -2 & 0 & 0 & 0 & 0 & 0 & 0 & 0 & -1 & 30
& 0 & 0 & -10 & 0 & 20 & -70   \\

$({\bf \overline{10}},{\bf 1})$ & 1 & -2 & 0 & 0 & 0 & 0 & 0 & 0 & 0 & -1 & 30
& 0 & 0 & -10 & 0 & 20 & -70   \\
\hline

\multicolumn{18}{|c|}{Additional chiral and non-chiral Matter}\\
\hline

\end{tabular}
\caption{The particle spectrum in the
observable sector in the Model FSU(5)-III, with 
the four global $U(1)$s from
the Green-Schwarz mechanism.  The $\star'd$ representations
indicate vector-like matter. }
\label{FSU5-3-spectrum}
\end{center}
\end{table}

The $U(1)_X$ gauge symmetry is
\begin{equation}
U(1)_X= \frac{1}{2}(U(1)_a -5U(1)_b +5U(1)_c -5U(1)_d +5U(1)_e
+5U(1)_f + 5U(1)_g +5 U(1)_h)~.~\,
\end{equation}
The four global $U(1)$'s are
\begin{eqnarray}
U(1)_1&=& -15U(1)_a + 3 U(1)_c + 27 U(1)_e - 27 U(1)_f - 21U(1)_g + 21 U(1)_h
~,~\, \nonumber \\
U(1)_2&=& - U(1)_e + U(1)_f~,~\, \nonumber \\
U(1)_3&=& U(1)_b - U(1)_d~,~\, \nonumber \\
U(1)_4&=& 5U(1)_a -3 U(1)_b - U(1)_c + 3U(1)_d + 7U(1)_g -7
U(1)_h~.~\,
\end{eqnarray}
And the other massless $U(1)$'s are:
\begin{eqnarray}
U(1)_U &=& U(1)_e + U(1)_f -U(1)_g - U(1)_h~,~\,  \nonumber \\
U(1)_V &=& -10 U(1)_a -50U(1)_c + 13 U(1)_e + 13 U(1)_f +13 U(1)_g + 13 U(1)_h ~,~\, 
\nonumber \\
U(1)_W &=& 35U(1)_a -7 U(1)_c -13 U(1)_g +13 U(1)_h~.~\,
\end{eqnarray}

\newpage

\subsection{Model FSU(5)-IV}

We present the D6-brane configurations and intersection numbers for
the Model FSU(5)-IV on Type IIA $\mathbf{T^6/(\Z_2\times \Z_2)}$
 orientifold in Tables \ref{FSU5-4} and \ref{FSU5-4C},  and its
particle spectrum in the observable sector in
Table \ref{FSU5-4-spectrum}.

\begin{table}[h]
\renewcommand{\arraystretch}{.9}
\begin{center}
\footnotesize
\begin{tabular}{|@{}c@{}|c||@{}c@{}c@{}c@{}||@{}c@{}|@{}c@{}||c|@{}c@{}|@{}c@{}|@{}c@{}|
c|c|c|c|@{}c@{}|@{}c@{}|c|c|} \hline

stk & $N$ & ($n_1$,$l_1$) & ($n_2$,$l_2$) & ($n_3$,$l_3$) & A & S
& $b$ & $b'$ & $c$ & $c'$ & $d$ & $d'$ & $e$ & $e'$ & $f$ & $f'$ &
$g$ & $g'$     \\ \hline \hline

$a$ & 10 & ( 1, 3) & ( 1, 1) & ( 0,-1) & 2 & -2 & -3 & 0(1) & 0(2)
& 0(3) & 0 & 0 & 24 & 24 & 12 & -6 & 6 & -3   \\ \hline

$b$ & 2 & ( 1,-3) & ( 0,-1) & (-1, 1) & -2 & 2 & - & - & -3 & 0(1)
& 2 & -1 & -15 & -60 & -12 & 6 & 12 & 12    \\ \hline

$c$ & 2 & ( 1, 3) & (-1, 1) & (-1, 0) & 2 & -2 & - & - & - & - & 4
& 4 & 0 & 0 & 18 & 36 & -18 & -9   \\  \hline

$d$ & 2 & ( 1, 1) & (-1,-3) & ( 0, 1) & 2 & -2 & - & - & - & - & -
& - & 84 & -36 & 6 & 0 & 15 & -12    \\  \hline

$e$ & 2 & (-5, 9) & (-5,-3) & ( 1, 0) & -30 & 30 & - & - & - & - &
- & - & - & - & -486 & -324 & 162 & 243   \\  \hline

$f$ & 2 & ( 2, 0) & (-1, 3) & (-1,-3) & 0 & 0 & - & - & - & - & -
& - & - & - & - & - & 0 & -162    \\ \hline

$g$ & 2 & ( 1,-9) & (-1, 0) & (-1,-3) & -6 & 6 & - & - & - & - & -
& - & - & - & - & - & - & -   \\ \hline

$h$ & 2 & ( 1,-7) & ( 0, 1) & ( 7,-3) & 0 & 0 & - & - & - & - & -
& - & - & - & - & - & - & -   \\ \hline

$i$ & 2 & ( 0, 2) & ( 4,-3) & ( 3,-4) & 0 & 0 & - & - & - & - & -
& - & - & - & - & - & - & -    \\ \hline

$j$ & 2 & ( 1,-3) & (-1, 0) & (-1,-1) & -2 & 2 & - & - & - & - & -
& - & - & - & - & - & - & -   \\ \hline

$k$ & 2 & ( 0, 2) & (-3,-1) & ( 1, 3) & 0 & 0 & - & - & - & - & -
& - & - & - & - & - & - & -    \\ \hline

$O6^1$ & 10 & ( 2, 0) & ( 1, 0) & ( 1, 0) & - & - & - & - & - & -
& - & - & - & - & - & - & - & -  \\ \hline

$O6^2$ & 10 & ( 2, 0) & ( 0,-1) & ( 0, 1) & - & - & - & - & - & -
& - & - & - & - & - & - & - & -  \\ \hline

$O6^3$ & 8 & ( 0,-2) & ( 1, 0) & ( 0, 1) & - & - & - & - & - & - &
- & - & - & - & - & - & - & -  \\ \hline

$O6^4$ & 2 & ( 0,-2) & ( 0, 1) & ( 1, 0) & - & - & - & - & - & - &
- & - & - & - & - & - & - & -  \\  \hline
\end{tabular}
\caption{D6-brane configurations and intersection numbers (Part 1)
for the Model FSU(5)-IV on Type IIA $\mathbf{T^6/(\Z_2\times
\Z_2)}$ orientifold. The complete gauge symmetry is $U(5) \times
U(1)^{10} \times USp(10)^2 \times USp(8) \times USp(2)$, and the
complex structure parameters are $\chi_1 =2/3, \;\; \chi_2=1$, and
$\chi_3 =1$. To satisfy the RR tadpole cancellation conditions, we
choose $h_0=-2(3q+8)$, $a=16$, and $m=8$.} 
\label{FSU5-4}
\end{center}
\end{table}

\begin{table}[h]
\renewcommand{\arraystretch}{.9}
\begin{center}
\footnotesize
\begin{tabular}{|@{}c@{}|c||@{}c@{}c@{}c@{}||@{}c@{}|@{}c@{}||
@{}c@{}|@{}c@{}|@{}c@{}|@{}c@{}|@{}c@{}|@{}c@{}|@{}c@{}|@{}c@{}|@{}c@{}|@{}c@{}|@{}c@{}|@{}c@{}|}
\hline

stk & $N$ & ($n_1$,$l_1$) & ($n_2$,$l_2$) & ($n_3$,$l_3$) & $h$ &
$h'$ & $i$ & $i'$ & $j$ & $j'$ & $k$ & $k'$ & $O6^1$ & $O6^2$ &
$O6^3$ & $O6^4$     \\ \hline \hline

$a$ & 10 & ( 1, 3) & ( 1, 1) & ( 0,-1) & -35 & -14 & -21 & 3 & 3 &
0 & 2 & -4 & 3 & 0 & 0 & -1  \\ \hline

$b$ & 2 & ( 1,-3) & ( 0,-1) & (-1, 1) & 0 & 0 & 4 & 28 & 0 & 0 &
12 & 6 & -3 & 0 & 1 & -3   \\ \hline

$c$ & 2 & ( 1, 3) & (-1, 1) & (-1, 0) & 15 & -6 & -4 & -28 & -3 &
0 & -12 & -6 & 0 & 3 & -1 & 0  \\  \hline

$d$ & 2 & ( 1, 1) & (-1,-3) & ( 0, 1) & -28 & -21 & -45 & 27 & 6 &
-3 & 8 & -10 & 3 & 0 & 3 & -1    \\  \hline

$e$ & 2 & (-5, 9) & (-5,-3) & ( 1, 0) & 195 & -330 & 540 & -60 & 9
& -36 & 60 & 210 & 0 & -45 & 15 & 0   \\  \hline

$f$ & 2 & ( 2, 0) & (-1, 3) & (-1,-3) & 168 & 126 & -234 & 150 &
18 & -36 & 0 & -96 & 0 & 0 & -6 & 6   \\ \hline

$g$ & 2 & ( 1,-9) & (-1, 0) & (-1,-3) & -24 & 144 & 39 & 15 & 0 &
0 & 0 & 6 & 0 & -9 & 0 & 3  \\ \hline

$h$ & 2 & ( 1,-7) & ( 0, 1) & ( 7,-3) & - & - & 76 & 0 & -20 & 20
& 72 & 54 & -21 & 0 & 7 & 0   \\ \hline

$i$ & 2 & ( 0, 2) & ( 4,-3) & ( 3,-4) & - & - & - & - & -21 & -3 &
0 & 0 & -24 & 24 & 0 & 0    \\ \hline

$j$ & 2 & ( 1,-3) & (-1, 0) & (-1,-1) & - & - & - & - & - & - & -2
& 4 & 0 & -3 & 0 & 1  \\ \hline

$k$ & 2 & ( 0, 2) & (-3,-1) & ( 1, 3) & - & - & - & - & - & - & -
& - & 6 & -6 & 0 & 0   \\ \hline

$O6^1$ & 10 & ( 2, 0) & ( 1, 0) & ( 1, 0) & - & - & - & - & - & -
& - & - & - & - & - & - \\ \hline

$O6^2$ & 10 & ( 2, 0) & ( 0,-1) & ( 0, 1) & - & - & - & - & - & -
& - & - & - & - & - & - \\ \hline

$O6^3$ & 8 & ( 0,-2) & ( 1, 0) & ( 0, 1) & - & - & - & - & - & - &
- & - & - & - & - & - \\ \hline

$O6^4$ & 2 & ( 0,-2) & ( 0, 1) & ( 1, 0) & - & - & - & - & - & - &
- & - & - & - & - & -  \\  \hline
\end{tabular}
\caption{D6-brane configurations and intersection numbers (Part 2)
for the Model FSU(5)-IV on Type IIA $\mathbf{T^6/(\Z_2\times
\Z_2)}$ orientifold.} 
\label{FSU5-4C}
\end{center}
\end{table}

\begin{table}[h]
\renewcommand{\arraystretch}{1}
\begin{center}
\footnotesize
\begin{tabular}{|@{}c@{}||@{}c@{}||@{}c@{}||@{}c@{}|@{}c@{}|@{}c@{}|
@{}c@{}||@{}c@{}|@{}c@{}|c|} \hline

Rep. & Multi.  & $U(1)_X$ & $U(1)_1$ & $U(1)_2$ & $U(1)_3$ &
$U(1)_4$ & $U(1)_U$ & $U(1)_V$ & $\; \cdots$   \\
\hline \hline

$({\bf 10},{\bf 1})$ & 3 & 1 & 0 & 0 & -20 & 60 & 0 & 0 & $\cdots$   \\

$({\bf \bar{5}}_a,{\bf 1}_b)$ & 3 & -3 & 0 & 2 & 10 & -36 & -10 & 125 & $\cdots$  \\

$({\bf \bar{1}}_b,{\bf 1}_c)$ & 3 & 5 & 6 & -4 & 0 & 6 & 11 & -205 & $\cdots$
\\ \hline \hline

$({\bf 10},{\bf 1})$ & 1 & 1 & 0 & 0 & -20 & 60 & 0 & 0 & $\cdots$   \\

$({\bf \overline{10}},{\bf 1})$ & 1 & -1 & 0 & 0 & 20 & -60 & 0 & 0 & $\cdots$  \\
\hline

$({\bf 5}_a,{\bf 1}_b)^\star$ & 1 & -2 & 0 & 2 & -10 & 24 & -10 & 125 & $\cdots$   \\

$({\bf \bar{5}}_a,{\bf \bar{1}}_b)$ & 1 & 2 & 0 & -2 & 10 & -24 & 10 & -125 &
$\cdots$ \\ \hline

$({\bf 1}_c,{\bf \bar{1}}_d)$ & 4 & 0 & 6 & -2 & 2 & -6 & 1 & -80 & $\cdots$  \\
\hline \hline

$({\bf \overline{15}},{\bf 1})$ & 2 & -1 & 0 & 0 & 20 & -60 & 0 & 0 & $\cdots$    \\

$({\bf \overline{10}},{\bf 1})$ & 1 & -1 & 0 & 0 & 20 & -60 & 0 & 0 & $\cdots$    \\
\hline

\multicolumn{10}{|c|}{Additional chiral and non-chiral Matter}\\
\hline

\end{tabular}

\caption{The particle spectrum in the
observable sector in the Model FSU(5)-IV, with 
the four global $U(1)$s from
the Green-Schwarz mechanism.  The $\star'd$ representations
indicate vector-like matter. }
\label{FSU5-4-spectrum}
\end{center}
\end{table}

The $U(1)_X$ gauge symmetry is
\begin{eqnarray}
U(1)_X &=& \frac{1}{2}(U(1)_a -5U(1)_b +5U(1)_c +5U(1)_d +5U(1)_e
+5U(1)_f - 5U(1)_g \nonumber \\
&& - 5 U(1)_h + 5U(1)_i -5U(1)_j - 5 U(1)_k)~.~\,
\end{eqnarray}

And the four global $U(1)$'s are
\begin{eqnarray}
U(1)_1&=& 6 U(1)_c -90U(1)_e -18 U(1)_g +48 U(1)_i -6 U(1)_j -12 U(1)_k~,~\, \nonumber \\
U(1)_2&=& 2 U(1)_b -2 U(1)_c +30 U(1)_e -12 U(1)_f + 14U(1)_h~,~\, \nonumber \\
U(1)_3&=& -10 U(1)_a -2 U(1)_d + 12U(1)_f + 6U(1)_g +2 U(1)_j~,~\, \nonumber \\
U(1)_4&=& 30 U(1)_a -6 U(1)_b +6 U(1)_d - 42U(1)_h -48 U(1)_i +12
U(1)_k~.~\, \;\;\;\;
\end{eqnarray}
There are six other massless  $U(1)$'s. As an example, we present two
of them:
\begin{eqnarray}
U(1)_U&=& -10 U(1)_b + U(1)_c - U(1)_e -2 U(1)_f +4 U(1)_g +2 U(1)_h +2U(1)_k
~,~\, \nonumber \\
U(1)_V&=& 125 U(1)_b -80 U(1)_c + 26 U(1)_e -85 U(1)_h +47 U(1)_i- 47U(1)_k
~.~\,
\end{eqnarray}

%%%%%%%%%%%%%%%%%%%%%%%%%%%%%%%%%%%%%%%%%%%%%%%%%%%%%%%%%%%%%%%%
% XXXXXXXXXXXXXXXXXXXXXXXXXXXXXXXXXXXXXXXXXXXXXXXXXXXXXXXXX
%%%%%%%%%%%%%%%%%%%%%%%%%%%%%%%%%%%%%%%%%%%%%%%%%%%%%%%%%%%%%%%%

\section{Discussion and Conclusions}

On Type IIA orientifolds
with flux compactifications in supersymmetric AdS vacua,
we for the first time constructed the exact three-family
$SU(5)$ models.
In these models, we have three ${\bf 10}$ representations,
and obtain three ${\bf \overline{5}}$ representations
after the additional gauge symmetry breaking via supersymmetry
preserving Higgs mechanism. 
So, there are exact three families of the SM fermions,
and no chiral exotic particles that
are charged under $SU(5)$.
In addition, we can break the $SU(5)$ gauge symmetry 
 down to the SM gauge symmetry via D6-brane
splitting, and solve the doublet-triplet splitting problem.
If the extra one (or several) pair(s) of Higgs doublets and
adjoint particles obtain GUT/string scale masses
via high-dimensional operators, we only have the MSSM in
the observable sector below the GUT scale. 
Choosing suitable grand unified gauge coupling by adjusting the string scale,
we can explain the observed low energy gauge
couplings via RGE running. However,
 how to generate the up-type quark Yukawa couplings, which are
forbidden by the global $U(1)$ symmetry,
deserves further study.

Furthermore, we considered the flipped $SU(5)$ models.
In order to have at least one pair of Higgs
fields ${\bf 10}$ and ${\bf \overline{10}}$, we must have
the symmetric representations, and then the net number
of ${\bf \overline{5}}$ and ${\bf 5}$ can not be three
if the net number of ${\bf 10}$ and ${\bf \overline{10}}$ is three
due to the non-abelian anomaly free condition.
We constructed the first model with three
${\bf 10}$ representations, and the
first model where all the Yukawa couplings are allowed by
the global $U(1)$ symmetries.

\section*{Acknowledgments}

T.L. would like to thank R.~Blumenhagen for helpful discussions.
The research of T.L. was supported by DOE grant DE-FG02-96ER40959,
and the research of D.V.N. was supported by DOE grant
DE-FG03-95-Er-40917.

\end{document}